\documentclass[preprint,aps,floats,nofootinbib,amssymb,superscriptaddress]{revtex4}
\usepackage[sort&compress]{natbib}
\usepackage{epsf,epsfig}
\usepackage{amsmath}
\usepackage{hyperref}
\usepackage{subfigure}
\usepackage{graphicx}
\usepackage{color}
\usepackage{mathrsfs}

\newcommand{\be}{\begin{equation}}
\newcommand{\ee}{\end{equation}}
\newcommand{\bea}{\begin{eqnarray}}
\newcommand{\eea}{\end{eqnarray}}
\newcommand{\nn}{\nonumber}

\newcommand{\ba}{\begin{array}}
\newcommand{\ea}{\end{array}}
\newcommand{\bi}{\begin{itemize}}
\newcommand{\ei}{\end{itemize}}

\newcommand{\wplus}{W^{+}}
\newcommand{\wminus}{W^{-}}

\newcommand{\edote}[2]{ (\epsilon_{#1} \cdot \epsilon_{#2} ) }
\newcommand{\pdote}[2]{ (p_{#1} \cdot \epsilon_{#2} ) }

\newcommand{\eeee}[4]{ (\epsilon_{#1} \cdot \epsilon_{#2} ) (\epsilon_{#3} \cdot \epsilon_{#4}) }
\newcommand{\eepepe}[6]{ (\epsilon_{#1} \cdot \epsilon_{#2} ) (p_{#3} \cdot \epsilon_{#4} )(p_{#5} \cdot \epsilon_{#6}) }

\newcommand{\fracp}[2]{ \left( \frac{#1}{#2} \right)}

\begin{document}

%-----------------------------------
% Preprint numbers
%-----------------------------------
\preprint{ICCUB-12-480}
\preprint{UB-ECM-PF-84/12}

%-----------------------------------
% Title
%-----------------------------------
\title{\vspace*{.75in} Longitudinal $WW$ scattering in light of the ``Higgs boson" discovery}

%-----------------------------------
% Authors
%-----------------------------------
\author{Dom\`enec Espriu}\affiliation{Departament d'Estructura i Constituents de la Mat\`eria,
Institut de Ci\`encies del Cosmos (ICCUB), \\
Universitat de Barcelona, Mart\'i Franqu\`es 1, 08028 Barcelona, Spain}
\author{Brian Yencho}\affiliation{Departament d'Estructura i Constituents de la Mat\`eria,
Institut de Ci\`encies del Cosmos (ICCUB), \\
Universitat de Barcelona, Mart\'i Franqu\`es 1, 08028 Barcelona, Spain}

\thispagestyle{empty}

%-----------------------------------
% Abstract
%-----------------------------------

\begin{abstract}
$WW$ scattering is dominated at high energies by their longitudinal components, which are the most sensitive to the nature of the electroweak symmetry breaking. Prior to the discovery at the LHC of a Higgs-like particle, unitarization tools were extensively used to show that, in the absence of a light Higgs boson, new resonances resulting from the would-be strongly interacting electroweak sector would appear, and furthermore these techniques would approximately predict their masses, widths, and signal strengths.  With the discovery of a Higgs-like particle now firmly established, we reinvestigate these techniques assuming this particle couples exactly as in the SM, but still being open to the possibility of an extended symmetry breaking sector.  While the SM itself is free from problems with perturbative unitarity in the electroweak sector, ``anomalous" self-couplings of the vector bosons -- low-energy remnants of such higher-energy symmetry breaking sectors -- are easily shown to reintroduce them.  We demonstrate how new resonances should still appear in the scattering of electroweak vector bosons after imposing constraints from unitarity, and we discuss their ability to be probed with current and future LHC data.
\end{abstract}

\maketitle

%%%%%%%%%%%%%%%%%%%%%%%%%%%%
\section{Introduction} \label{sec:intro}
%%%%%%%%%%%%%%%%%%%%%%%%%%%% 
There appears to be growing evidence that the particle discovered by the LHC experiments~\cite{atlas,cms} is
quite close in properties to what is expected from the Higgs particle in 
the standard model. CMS reports tentative, yet suggestive, evidence~\cite{spin} that a $J^P= 0^+$ is clearly 
favored in front of $J^P=0^-$. As it is well known, $J=1$ is excluded due to the Landau-Yang theorem~\cite{yang} 
and the fact that the decay to two photons has been seen well above the 5$\sigma$ level~\cite{twophotons}. 
Although the distinction between $J=2$ and $J=0$ is still not possible with the 
amount of data available at the moment, the first possibility is certainly disfavored theoretically.

It is also reported that an additional standard model (SM) Higgs boson is excluded at present at the 95\% level in the range $\sim$130--600~GeV \cite{exclusion}. Furthermore no signal of any additional vector or scalar resonances has been seen in the data 
currently available. This absence of new resonances together with the results on spin-parity already available, and the
fact that most couplings so far measured are within the experimental errors comparable with the 
standard model\footnote{The coupling to two photons is actually slightly  off by about 1.5 $\sigma$ for CMS and by 
about 1.8 $\sigma$ for ATLAS with respect to its SM value according to the latest available data 
at the time of writing this paper~\cite{twophotons}}, would  strongly suggest that the simplest realization of the electroweak symmetry breaking sector in the standard model (EWSBS)
is strongly favored and there is no compelling reason to expect new particles associated with the EWSBS anytime soon.

This may be jumping too hastily to conclusions, however. Let us examine a bit critically the statement 
that ``an additional standard model Higgs boson is excluded in the $\sim$130--600 GeV range''. 
Even in two-Higgs-doublet models~\cite{twodoublets} it is well known~\cite{ec} that only a combination 
of the two $0^+$ scalars involved has Higgs-like couplings and, in particular, only this combination 
exhibits the property of nondecoupling characteristic of spontaneous symmetry breaking. The other additional spin zero 
states couple in a model-dependent way and they cannot really be rigorously excluded yet. Likewise in composite models
where new vector resonances would be present the couplings are model dependent, although 
qualitatively statements concerning their magnitude can be made in many cases.

In this work we would like to analyze critically the consequences that can be drawn from the apparent 
absence of new resonances in the range of energies explored so far. We shall have in 
mind a composite Higgs scenario (like the ones proposed in Ref.~\cite{composite}) without needing to commit 
ourselves at this point to any particular model. We would like to understand whether the apparent 
absence of new resonances really means that no new states exist or simply that the signal 
due to them may at this point be well below the present experimental sensitivity and, if this is the case, 
what would be the expected strength of the signal. We shall make use of the technique of 
effective Lagrangians where the information on states with a mass $m \gg M_H \simeq 125$ GeV
can be encoded in some effective coefficients. Consequently we shall therefore be able to place new bounds on these 
effective coefficients that can be compared with the limits on anomalous $W$ and $Z$ couplings~\cite{couplings} already 
derived from the early LHC data for $WW$ or $ZZ$ cross sections. It is obvious, however, that the sensitivity 
on possible departures of these anomalous couplings with respect to their standard model values is 
still poor.

The techniques that we shall use rely on analyticity and unitarity; we shall make intensive use of the inverse amplitude
method (IAM), amply used in hadronic physics~\cite{iam} and quite useful in heavy Higgs models~\cite{heavyhiggs} 
(now seemingly ruled out). We shall adapt the technique to allow the inclusion of a light resonance---the Higgs boson.
For our analysis we shall need at some point the full one-loop correction to $WW$ scattering in the standard model. 
Unfortunately this is a rather involved calculation that is available in full only numerically~\cite{full}, and thus very 
inconvenient for unitarization techniques. We have circumvented this problem by restricting ourselves to longitudinal $W$  
scattering and making partial use of the equivalence theorem~\cite{et}; in fact for the real part
of the one-loop correction only. Other than that, the external $W$ are 
dealt with exactly. The reason not to use the equivalence theorem from the very beginning is that
at the moderate energies involved in our considerations the replacement of longitudinal $W$'s by the equivalent
Goldstone bosons has large corrections~\cite{esma} and the ensuing discussion would be unreliable.

The use of unitarity methods forces upon us an additional approximation albeit not an important one. 
We shall neglect throughout electromagnetic 
corrections as they do not lead to convergent partial wave amplitudes due to their long-range character. If desired, 
electromagnetic corrections could be reintroduced perturbatively. Neglecting electromagnetism brings for us a
bonus: we can use the isospin formalism that simplifies considerably the analysis. Another subtle point that will be discussed in 
detail is the use of crossing symmetry involving longitudinal $W$ scattering.   

We have presented our results in the following order. In Sec.~\ref{sec:echl} we review the electroweak chiral Lagrangian and how it is 
modified by the inclusion of a light scalar Higgs-like degree of freedom. Sec.~\ref{sec:isospin} is devoted to the introduction
of the different (weak-) isospin amplitudes and how the usual techniques have to be modified when considering the scattering
of physical longitudinal $W,Z$ bosons as opposed to Goldstone bosons as per the equivalence theorem. The inverse amplitude
method is discussed in detail in Sec.~\ref{sec:iam}, with a particular emphasis on the restoration
of unitarity. In Sec.~\ref{sec:calculation}, we present the details of our treatment of the relevant amplitudes.
In Sec.~\ref{sec:resonances}, we determine the resonances that different values
of the higher-order operators coefficients (beyond the SM) generate. In this section we will also compare the results 
obtained by use of unitarization techniques to the predictions of the minimal SM and also to the existing results for a heavy
Higgs boson (now excluded) for comparison. In Sec.~\ref{sec:xsec}, we will compute the corresponding cross 
sections and see what signals can be expected 
for additional resonances present in composite Higgs scenarios.

%%%%%%%%%%%%%%%%%%%%%%%%%%%%
\section{Electroweak Chiral Lagrangian} \label{sec:echl}
%%%%%%%%%%%%%%%%%%%%%%%%%%%% 
The effective Lagrangian that contains the light degrees of freedom in the standard model, other 
than the Higgs particle, relevant at scales below any new thresholds, is
\bea
\label{eq:l1}
\mathcal{L}^{\rm eff} & = & - \frac{1}{2} {\rm Tr} W_{\mu\nu} W^{\mu\nu} - \frac{1}{4} {\rm Tr} B_{\mu\nu} B^{\mu\nu}
+ \sum_{i=0,13} \mathcal{L}_{i} + \mathcal{L}_{\rm GF} + \mathcal{L}_{\rm FP} \\ \nn
& &
+ \frac{v^{2}}{4} {\rm Tr} D_{\mu}U^{\dagger} D^{\mu} U
\eea
\noindent where the electroweak-theory Goldstone bosons are given in the nonlinear representation
\be
U=\exp\left(i \frac{ (\vec{w} \cdot \vec{\tau})}{v} \right) \, ,
\ee
\noindent where $v\approx246$~GeV is the SM vacuum expectation value, $\tau_{i}$ are the Pauli matices, and $w_{i}$ are the Goldstone fields, which are related to the charged basis in the usual way: $w^{\pm} = (w_{1} \mp i w_{2})/\sqrt{2}$ and $z \equiv w^{0} = w_{3}$.  The covariant derivative of $U$ is then defined as
\be
D_{\mu} U =  \partial_{\mu}U + \frac{1}{2} i g W_{\mu}^{i} \tau^{i} U - \frac{1}{2} i g' B_{\mu}^{i} U \tau^{3}.
\ee
The $\mathcal{L}_{i}$ depend on unknown coefficients $a_{i}$ and are given 
in Appendix~\ref{sec:appendix_operators}.  Inspired by the nonlinear realization of the SM, we can 
add the Higgs field by writing
\bea
\label{eq:l2}
\mathcal{L}^{\rm eff}  & = & - \frac{1}{2} {\rm Tr} W_{\mu\nu} W^{\mu\nu} - \frac{1}{4} {\rm Tr} B_{\mu\nu} B^{\mu\nu}
+ \sum_{i=0,13} \mathcal{L}_{i} + \mathcal{L}_{\rm GF} + \mathcal{L}_{\rm FP} \\ \nn
& &
+ \left(1+\frac{h}{v} \right)^{2} \frac{v^{2}}{4} {\rm Tr} D_{\mu}U^{\dagger} D^{\mu} U
+ \frac{1}{2} \left( \partial_{\mu}h \right)^{2}
- \frac{1}{2} M_{H}^{2} h^{2}
\eea
This reproduces the SM interactions of the Higgs boson with the electroweak bosons in a gauge-invariant way. Recall that   
$h$ is a gauge singlet.  When the above coefficients $a_{i}$ are all taken to be zero and the appropriate 
tadpole and $\mathcal{O}(h^{3})$ and higher terms are added (they have been omitted in the previous expression), 
this is simply the nonlinear realization of the EWSBS.

If we wish to consider extensions or modifications of the EWSBS we can modify the preceding Lagrangian in
two ways. First, it may well be that the ``Higgs boson'' couples in a way that is different to the precise prescription
given in (\ref{eq:l2}). For instance we could write
\be
\label{eq:l21}
\left(1+f\fracp{h}{v} \right)^{2} \frac{v^{2}}{4} {\rm Tr} D_{\mu}U^{\dagger} D^{\mu} U.
\ee
Gauge invariance poses no restrictions on the form of $f(\frac{h}{v})$. If one assumes that this function, once Taylor expanded, 
behaves  as $\sim \frac{h}{v}$ for small values of $h$, then the ``Higgs boson'' is to be interpreted as a vacuum fluctuation, but other 
couplings can still depart from their minimal SM values in vertices involving more legs. This  
situation would present itself if the ``Higgs boson'' itself participates in some strong dynamics, e.g. in the dynamical Higgs boson scenarios 
suggested in Ref.~\cite{composite}.

In addition, the existence of an alternative EWSBS would for sure imply new heavier degrees of freedom. Their contribution at low 
energies can be collected in the effective coefficients $a_i$ and it does not affect the operators of dimension four present
in (\ref{eq:l2}). The extended dynamics may actually affect the ``Higgs boson'' interaction as well, but this effect is already 
accounted for by the function  $f(\frac{h}{v})$ and eventually by allowing the dimensionless coefficients $a_i$ to be 
functions of $\frac{h}{v}$ too. This last modification is not presently relevant to us. See, however, Ref.~\cite{belen}
for a recent discussion on this point.

The Lagrangian (\ref{eq:l1}) was extensively used in the past in a scenario, now ruled out, where the Higgs particle 
was assumed to be very heavy~\cite{heavyhiggs} or even absent, such as in simple QCD-technicolorlike models~\cite{technicolor},
mostly discussed in the context of electroweak precision observables.
In these models the coefficients $a_i$ serve also the important purpose of absorbing divergences that appear when computing one-loop 
corrections from (\ref{eq:l1}). However, the dimension four pieces of (\ref{eq:l2})---with the SM Higgs boson
explicitly thrown in---constitute by themselves a renormalizable subset and no extra divergences appear. The $a_i$ 
coefficients are therefore finite. Yet, if the function  $f(\frac{h}{v})$ departs from its standard model value, 
renormalizability will be in general lost and the $a_i$ will be needed to render the calculations finite. In this work
we shall adopt the conservative point of view that the couplings of the particle observed at the LHC are identical
to the ones of the standard model\footnote{We are aware that composite models could actually modify the one- and two-Higgs coupling to the electroweak vector bosons.  Here we consider the worst possible case (from the point of view of detecting new physics) in which these couplings are in practice indistinguishable from their values in the minimal SM.  We are therefore only considering a special case for the purpose of illustration.  Recent discussions of the effects of deviations in the Higgs boson couplings on tree-level perturbative unitarity can be found in Ref.~\cite{notreeunitarity}} and that only the vector boson \textit{self-couplings} may be allowed to deviate.

The main purpose of this paper is to determine the influence of an extended EWSBS, parametrized by the coefficients 
$a_i$ of the higher-dimensional operators on the scattering of longitudinal $W$ and 
$Z$ and their unitarization and, in particular, in the expected pattern of additional scalar and vector resonances
once the existence of a light Higgs-like particle is taken into account.

%%%%%%%%%%%%%%%%%%%%%%%%%%%%
\section{Isospin Amplitudes} \label{sec:isospin}
%%%%%%%%%%%%%%%%%%%%%%%%%%%% 
As mentioned in the Introduction we shall ignore electromagnetic corrections, setting 
$c_{w} \to 1$ ($M_{z} \to M_{W} \equiv M$). This corresponds to an exactly custodially preserving theory and
we can then use standard isospin techniques to relate different amplitudes. Let us
define the scattering amplitudes of the longitudinally polarized $W^{a}$ bosons as
\be
A^{abcd} \equiv A \left( W^{a}_{L}W^{b}_{L} \to W^{c}_{L}W^{d}_{L} \right) \, .
\ee
In the high-energy limit, where by virtue of the equivalence theorem \cite{et} the corresponding Goldstone 
bosons replace the longitudinal parts of the $W^{a}_{L}$, these amplitudes satisfy the following well-known isospin relation
\be
\label{eq:isospin_goldstones}
A^{abcd}(p^{a},p^{b},p^{c},p^{d}) = \delta^{ab}\delta^{cd} A(s,t,u) + \delta^{ac}\delta^{bd} A(t,s,u) + 
\delta^{ad}\delta^{bc} A(u,t,s) \, ,
\ee
where crossing symmetry has been assumed and
where $s=(p^{a}+p^{b})^{2}$, $t=(p^{a}-p^{c})^{2}$, $u=(p^{a}-p^{d})^{2}$ are the usual Mandelstam variables.  
The fundamental amplitude is related to a subset of the possible amplitudes as
\be
A(s,t,u) = A^{1122} = A^{1133} = A^{2233} = \cdots \, ,
\ee
where the dots indicate the amplitudes with the pairs reversed.  When written in the more familiar 
charged basis, this is fully encapsulated in
\be
A(s,t,u) = A^{+-00} \, (=A^{+-33})  \, .
\ee
The three amplitudes in this basis that will be of interest are then
\bea
\label{eq:amplitudes_goldstones}
A^{+-00} & = & A(s,t,u) \\ \nn
A^{+-+-} & = & A(s,t,u) + A(t,s,u) \\ \nn
A^{++++} & = & A(t,s,u) + A(u,t,s)  \, .\nn
\eea
These can then be used to define the amplitudes $T_{I}$ with fixed values of isospin $I$ given by
\bea
\label{eq:fixed_isospin}
T_{0} & = & \langle 00\vert S\vert 00 \rangle  =   3 A^{+-00} +   A^{++++} \\ \nn
T_{1} & = & \langle 10\vert S\vert 10 \rangle  =   2 A^{+-+-} - 2 A^{+-00} - A^{++++} \\ \nn
T_{2} & = & \langle 20\vert S\vert 20 \rangle  =   A^{++++} \, . \nn 
\eea
In the subsequent discussion, we shall also need the amplitude for the process $W^+W^-\to hh$. Taking into account that the final state
is an isospin singlet and defining 
\be 
A^{+-}(s,t,u)= A(W^+W^-\to hh) \, ,
\ee
the projection of this amplitude to the $I=0$ channel gives 
\be
\label{eq:fixed_isospin_hh}
T_{H, 0}= \sqrt{3} A^{+-}(s,t,u).
%\tilde T_0= \sqrt{3} A^{+-}(s,t,u).
\ee

For much of our calculations, however, we will not be working in the high-energy limit in which the Goldstone
bosons originating from the $SU(2)_L \times SU(2)_R \to SU(2)_V$ breaking
can replace the longitudinal parts of the $W$ bosons.  We must then generalize the above results to account 
for an ambiguity introduced by the longitudinal polarization vector, which does not actually transform under
Lorentz transformations as a 4-vector.  When using their usual definitions, 
expressions involving the polarization vector $\epsilon_{L}^\mu$ can not be cast in terms of 
the Mandlestam variables $s$, $t$, and $u$ 
until \textit{after} an explicit reference frame has been chosen, as they can not themselves be written 
solely in terms of covariant quantities.  This renders these variables a rather inconvenient 
choice for the final expressions.  While these amplitudes still satisfy isospin and crossing 
symmetries, this is only clearly manifest, then, when they remain in terms of the external 4-momenta.  
Mindful of this fact, the generalized isospin relation should then be written as
\bea
\label{eq:isospin_general}
A^{abcd}(p^{a},p^{b},p^{c},p^{d})
& = & 
 \hspace{0.35cm}
\delta^{ab} \delta^{cd} A(p^{a},p^{b},p^{c},p^{d}) \\ \nn
& & + 
\delta^{ac} \delta^{bd} A(p^{a},-p^{c},-p^{b},p^{d}) \\ \nn
& & + 
\delta^{ad} \delta^{bc} A(p^{a},-p^{d},p^{c},-p^{b})  \, ,
\eea
with the corresponding amplitudes in Eq.~\ref{eq:amplitudes_goldstones} given by 
\bea
\label{eq:amplitudes_general}
A^{+-00} & = & A(p^{a},p^{b},p^{c},p^{d}) \\ \nn
A^{+-+-} & = & A(p^{a},p^{b},p^{c},p^{d}) + A(p^{a},-p^{c},-p^{b},p^{d}) \\ \nn
A^{++++} & = & A(p^{a},-p^{c},-p^{b},p^{d}) + A(p^{a},-p^{d},p^{c},-p^{b}) \, . \nn
\eea
The fixed-isospin amplitudes remain as in Eq.~\ref{eq:fixed_isospin}.

From here we can define the partial wave amplitudes for fixed isospin $I$ and total angular momentum $J$ as 
\be
t_{IJ} = \frac{1}{64\pi} \int_{-1}^{1} d(\cos\theta) P_{J}(\cos\theta) T_{I} \, ,
\ee
\noindent where the $P_{J}(x)$ are the Legendre polynomials and the $T_I$ amplitudes have been defined in (\ref{eq:fixed_isospin}).  We will concern ourselves with only the lowest nonzero partial wave amplitude in each isospin channel: $t_{00}$, $t_{11}$, and $t_{20}$.  These will be referred to as the scalar/isoscalar, vector/isovector, and isotensor amplitudes.  Partial wave unitarity requires these amplitudes to satisfy $|t_{IJ}|<1$ in the high-energy limit.  However, for nonzero values of the $O(p^4)$ coefficients $a_i$ this perturbative expansion gives a nonunitary behavior of the partial wave amplitudes  for large values of $s$. In order restore unitarity and, in doing so, extract information on higher resonances, 
the partial wave amplitudes have to be unitarized.

%%%%%%%%%%%%%%%%%%%%%%%%%%%%
\section{Inverse Amplitude Method} \label{sec:iam}
%%%%%%%%%%%%%%%%%%%%%%%%%%%% 
Nonrenormalizable models typically produce scattering amplitudes that grow with the scattering energy too fast, 
breaking the unitarity bounds~\cite{bounds} at some point or other. Chiral
descriptions of QCD~\cite{GL} are archetypal examples of this behavior and unitarization techniques have to be used
to recover unitarity. A convenient way to obtain unitary amplitudes is provided by the inverse amplitude 
method~\cite{iam}. This is not the place to provide a review of its justification and limitations, but suffice
only to say that when the physical value of the pion decay constant $f_\pi$ and the $O(p^4)$ low-energy 
coefficients $L_i$ (as defined e.g. in \cite{GL}, the counterpart of the $a_i$ in strong interactions) 
are inserted in the chiral Lagrangian and the IAM method is used,
the validity of the chiral expansion is considerably extended and one is able to reproduce the $\rho$ meson pole as 
well as many other properties of low-energy QCD~\cite{iam}.

Let us consider an effective theory model whose amplitudes admit a perturbative expansion. The expansion parameter
could be the momentum (normalized by some reference mass) or simply an expansion in some coupling constant. Let
$t_{IJ}$ be one such amplitude describing some elastic process. Then we expand in loops
\be\label{expansion}
t_{IJ}= t_{IJ}^{(0)}+t_{IJ}^{(2)}+t_{IJ}^{(4)}+ ... \, .
\ee
Then the IAM approximation to the full amplitude is
\be
\label{eq:t_iam}
t_{IJ} \approx \frac{t_{IJ}^{(0)}}{1-t_{IJ}^{(2)}/t_{IJ}^{(0)}},
\ee
which is identical to the [1,1] Pad\'e approximant to $t_{IJ}$ derived from (\ref{expansion}). 
The above expression obviously reproduces the first two orders of the perturbative expansion and, in addition, 
satisfies the necessary unitarity constraints, namely $|t_{IJ}|<1$ at high energies and 
\be
\label{eq:full_unitarity}
{\rm Im \,} t_{IJ}(s) = \sigma(s) |t_{IJ}(s)|^{2},
\ee
when the perturbative ingredients satisfy
\be
\label{eq:part_unitarity}
{\rm Im \,} t_{IJ}^{(2)}(s) = \sigma(s) |t_{IJ}^{(0)}(s)|^{2} \, ,
\ee
as they must from the optical theorem.  The formula Eq.~(\ref{eq:t_iam}) can be applied too in the inelastic case,
 i.e. when there is more than one channel. It will then satisfy, in our case, 
\be
\label{eq:full_unitarity_hh}
{\rm Im \,} t_{IJ}(s) = \sigma(s) |t_{IJ}(s)|^{2} + \sigma_{H}(s) |t_{H, IJ}(s)|^{2} \, ,
\ee
which is again guaranteed to the order we work by the IAM and the optical theorem for the lowest-order terms
\be
\label{eq:part_unitarity_hh}
{\rm Im \,} t_{IJ}^{(2)}(s) = \sigma(s) |t_{IJ}^{(0)}(s)|^{2} + \sigma_{H}(s) |t_{H, IJ}^{(0)}(s)|^{2},
\ee
where the phase space factors used here are given by
\be
\sigma(s) = \sqrt{1 - \frac{4 M^{2}}{s}} \hspace{1cm},\hspace{1cm}  \sigma_{H}(s) = \sqrt{1 - \frac{4 M_{H}^{2}}{s}}.
\ee
See e.g. \cite{twochannels} for a discussion on this point, as it is relevant to us due to the appearance of $hh$ states, 
along with intermediate $WW$ ones, at one loop.  This is only a concern, however, for the $I=0$ amplitudes.

If we examine the Lagrangian (\ref{eq:l1}) we see that it is formally identical to a gauged chiral effective Lagrangian
and it is therefore natural to use techniques that are known to work well in QCD at low energies, such as the IAM, in
the present context. This Lagrangian (\ref{eq:l1}) is the one describing the electroweak symmetry 
breaking sector at scales well below the Higgs mass and it has been thoroughly investigated using 
unitarization techniques in the past. 
We know from recent results that the Higgs particle is very likely light so this is not
a particularly relevant example anymore but let us first reexamine this case anyway with 
the only objective  to get an idea of the validity of the method.

The value of the higher-order coefficients for the standard model is obtained after matching $S$-matrix elements in the effective theory and in the standard model itself~\cite{matching}. For a heavy Higgs, the value of the relevant coefficients is shown in Appendix \ref{sec:appendix_operators}. We see that they are divergent (the modified minimal subtraction ($\overline{MS}$) scheme  is used throughout) since the theory that is left after removing the Higgs as a dynamical degree of freedom is nonrenormalizable.

Using these values, after unitarization, one is able by just using chiral perturbation theory techniques to reproduce the
pole that would correspond to a heavy SM Higgs with reasonable accuracy. This has been discussed in great detail in
the literature~\cite{heavyhiggs}, always in the context of the equivalence theorem. In these older studies the coefficient
$a_3$ does not play any role (it is absent if the equivalence theorem is used) and only $a_4$ and $a_5$ matter (see
Appendix \ref{sec:appendix_operators}). For the $I=J=0$ channel, the coefficients of higher-dimensional operators
always appear in the combination $7a_4 + 11a_5$ and therefore the scalar masses and widths obtained after unitarization
depend on this combination only. The equivalence theorem combined with the IAM procedure actually reproduces fairly well
a Higgs from around 500 GeV (width 45 GeV) up to $M_H\simeq 1500$ GeV (width $\simeq 1500$ GeV). At this point, the
widths become so large that the resonances ``melt''. Below $M_H\simeq 500$ GeV the method becomes progressively
unreliable due to the limitations of the equivalence theorem. Note also that vector resonances, i.e. poles for
$I=J=1$ in the second Riemann sheet, may also appear. This time 
the relevant combination of coupling constants is $a_4-2 a_5$ and, when present, their masses are characteristically higher
than the scalar ones.
It is characteristic of this analysis that the values of $a_4, a_5$ where scalar and vector resonances appear
are rather disjoint. See the last reference in Ref.~\cite{technicolor} for further details.   

We could also turn to a completely different case, namely the  minimal standard model with a light Higgs boson---and nothing 
else, i.e. no $O(p^4)$ coefficients, so the $a_i$ are all set to zero.  Of course, this a weakly coupled (and unitary) theory 
and perturbation theory should be an excellent guidance, but nothing prevents us from applying the machinery of the IAM nevertheless. 
The result of this exercise will be shown in Sec.~\ref{sec:resonances}.

The preceding discussion can be summarized by saying that the IAM reproduces the general features expected from the lightest resonances in strongly interacting models such as QCD, or extensions of the SM such as technicolor or models with a heavy Higgs boson, and also in weakly interacting theories, such as the minimal SM with a light Higgs boson. The limitations of the model derive from the accuracy in our knowledge of the different amplitudes entering the game (hence we have to abandon as much as possible the use of the equivalence theorem for a light Higgs boson) and in the validity of the approximations made in using the effective Lagrangians (\ref{eq:l2}) and (\ref{eq:l1}).  

The range of validity of the effective Lagrangian (\ref{eq:l2}) should thus be established. It is in principle valid down 
to arbitrarily low energies if we use longitudinal $W$'s and by doing so bypass the limitations of the 
equivalence theorem. In the high-energy range, it is in principle {\em perturbatively} 
valid until a resonance is encountered in a given channel but its 
validity can be largely extended by the unitarization process. This will allow us to use the bounds 
already available on additional resonances to constrain the higher-dimensional operators in (\ref{eq:l2}).
In any case the range can extend at most to $s\simeq  (4\pi v)^2 \simeq (3~{\rm TeV})^2$ as this is the natural parameter
in the momentum expansion.

%%%%%%%%%%%%%%%%%%%%%%%%%%%
\section{Calculation} \label{sec:calculation}
%%%%%%%%%%%%%%%%%%%%%%%%%%%%
We compute the tree-level contribution both from the lowest-order standard model terms and from the $O(p^4)$ 
coefficients in Eq.~\ref{eq:l2} exactly, i.e. without having to appeal to the equivalence theorem. The reason---previously
mentioned---is that in the standard model tree-level contribution, changes with respect to the simplest version of the 
equivalence theorem can be quite substantial for certain angles (i.e. certain values of the Mandelstam variable $t$). Indeed 
it was seen in \cite{esma} that using $W_L$ rather than the equivalent Goldstone boson $w$ makes a significant difference. 
In addition, we want to make sure that all
kinematic singularities are properly included at tree level. In the terms describing higher resonances (i.e. in
the $a_i$ coefficients, see Appendix \ref{sec:appendix_operators}) we shall only consider for the time being
 custodially preserving terms. In fact, we
will in the present analysis not consider $a_3$, as it turns out to make a relatively small contribution, and we will therefore concentrate on the contributions from the remaining two custodially preserving coefficients $a_4$ and $a_5$.

\begin{figure}[tb]
\centering
\subfigure[]{\includegraphics[clip,width=0.30\textwidth]{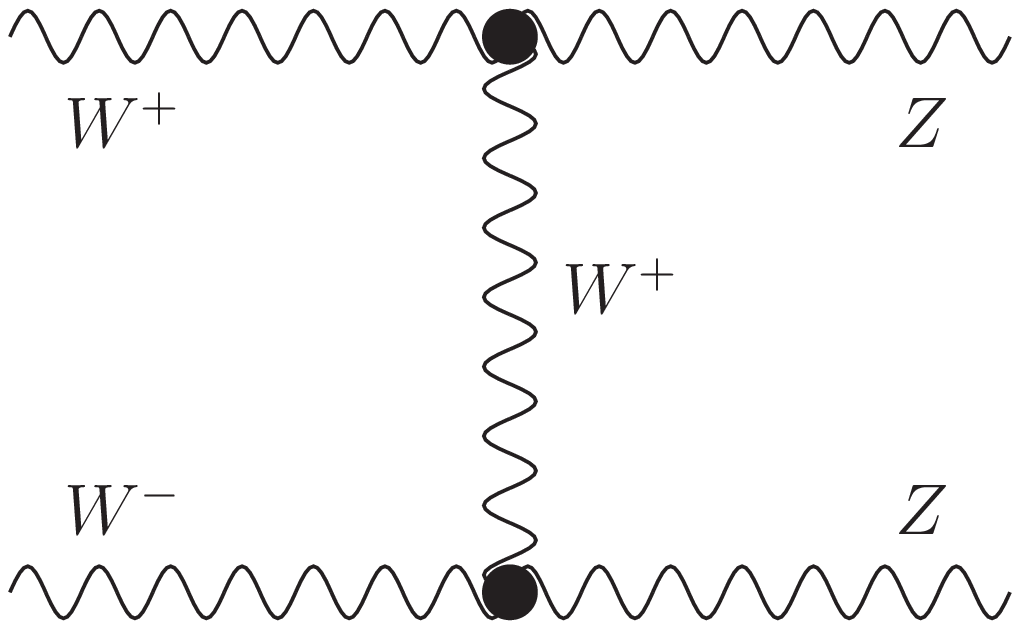}} \hspace{1cm}
\subfigure[]{\includegraphics[clip,width=0.30\textwidth]{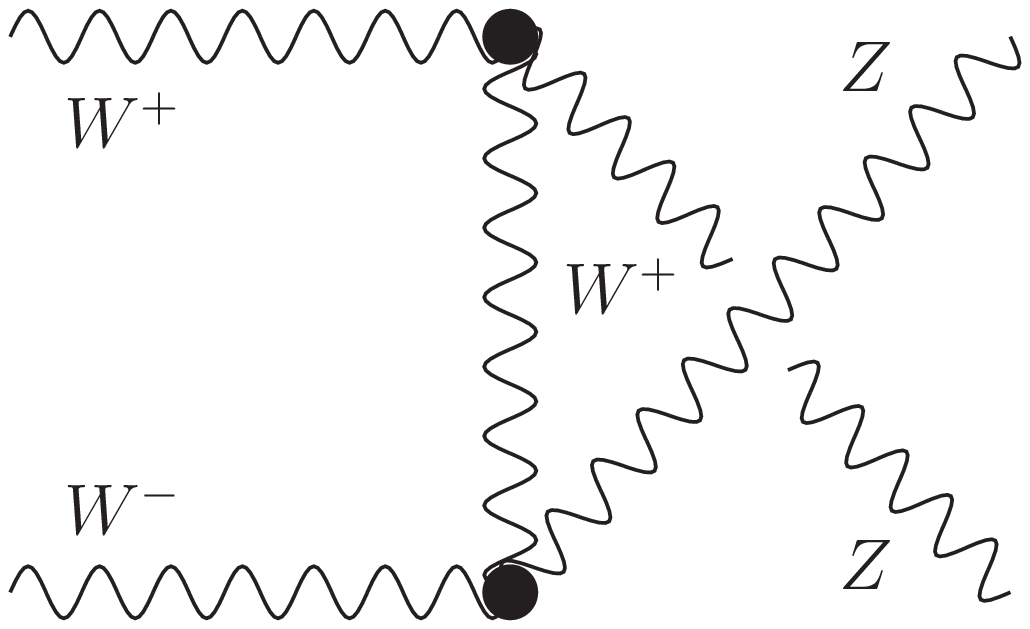}} \\
\subfigure[]{\includegraphics[clip,width=0.30\textwidth]{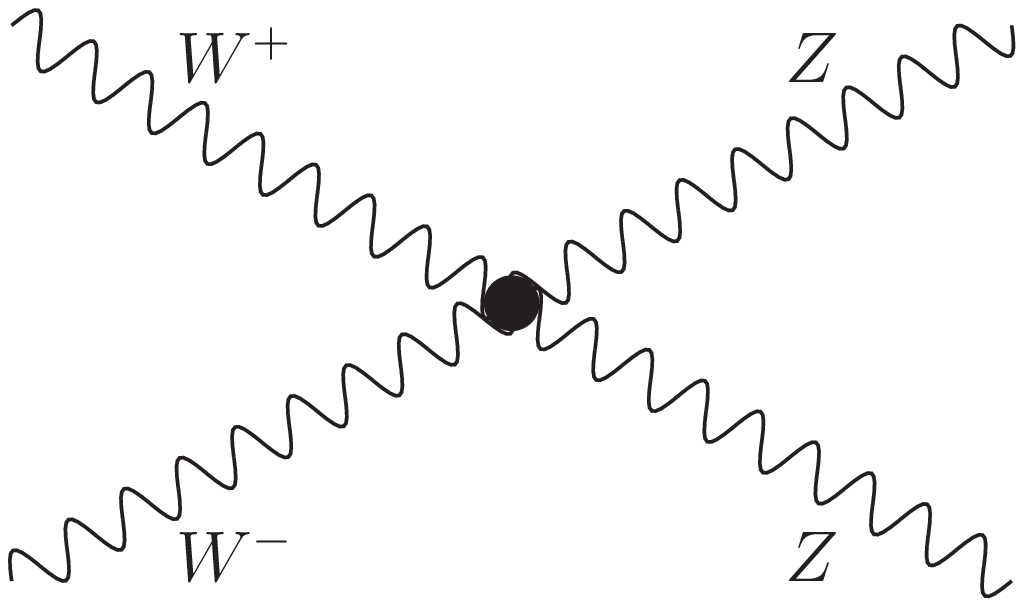}} \hspace{1cm}
\subfigure[]{\includegraphics[clip,width=0.30\textwidth]{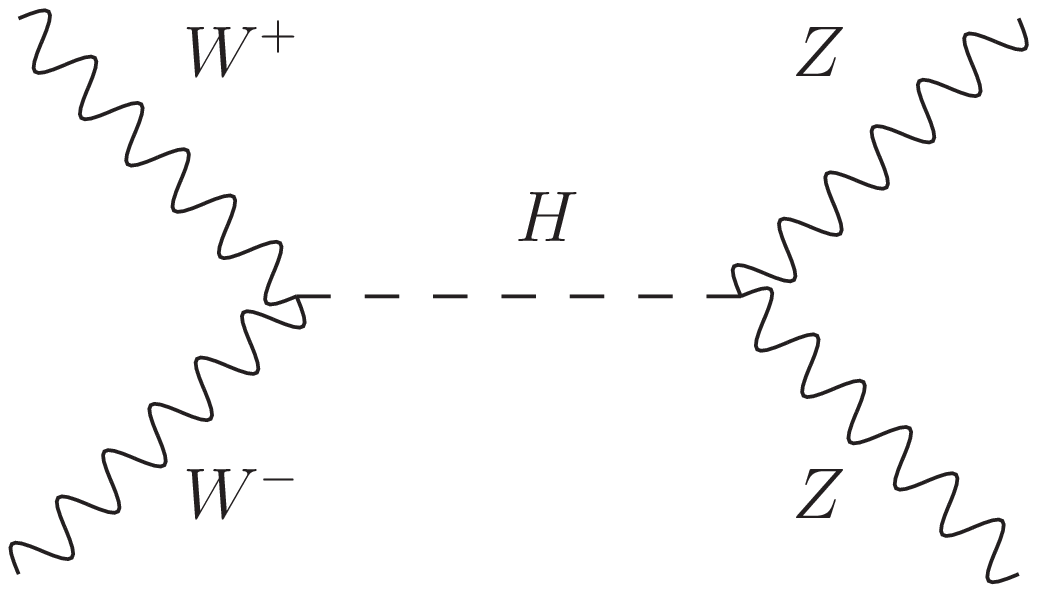}}
\caption{$\wplus \wminus \to ZZ$ scattering amplitudes at tree level.  Black dots indicate vertices modified in our calculation by the presence of nonzero coefficients $a_{i}$.}
\label{fig:amplitudes_tree}
\end{figure}

The calculation, then, involves the tree-level and $a_{i}$-dependent results for $A(W^{+}W^{-} \to Z Z)$ according 
to the diagrams of Figs.~\ref{fig:amplitudes_tree}(a)-(d).  The black dots indicate vertices 
which receive contributions from the ordinary Lagrangian terms as well as $\mathcal{L}_{i}$ terms and 
therefore depend on the $a_{i}$.  In the absence of $a_{3}$, only the quartic term in Fig.~\ref{fig:amplitudes_tree}(c) is modified. The full amplitude is given in Appendix~\ref{sec:appendix_amplitudes}.  

\begin{figure}[tb]
\centering
\subfigure[]{\includegraphics[clip,width=0.30\textwidth]{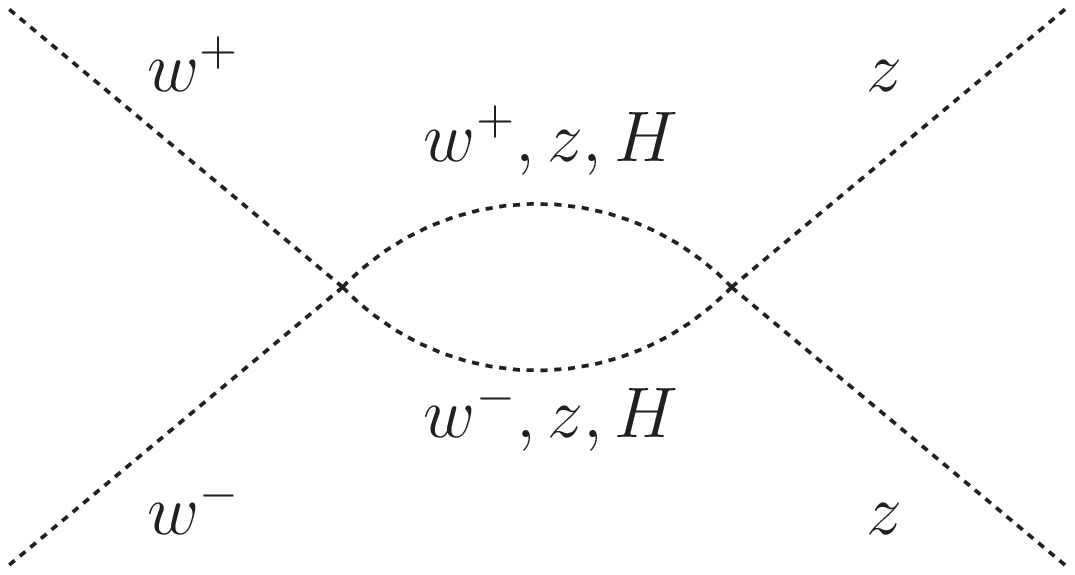}} \\
\subfigure[]{\includegraphics[clip,width=0.30\textwidth]{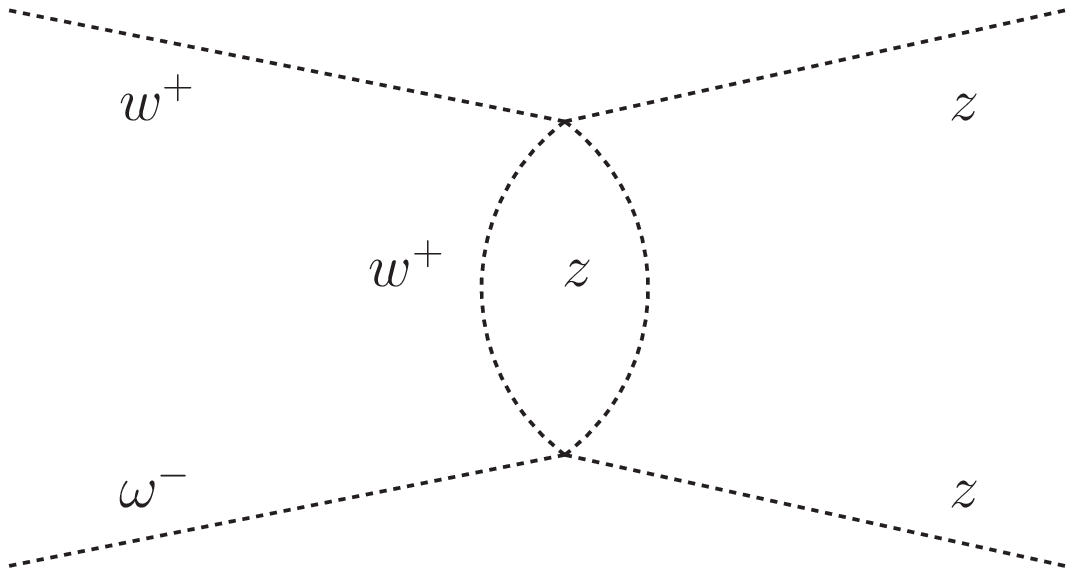}} \hspace{1cm}
\subfigure[]{\includegraphics[clip,width=0.30\textwidth]{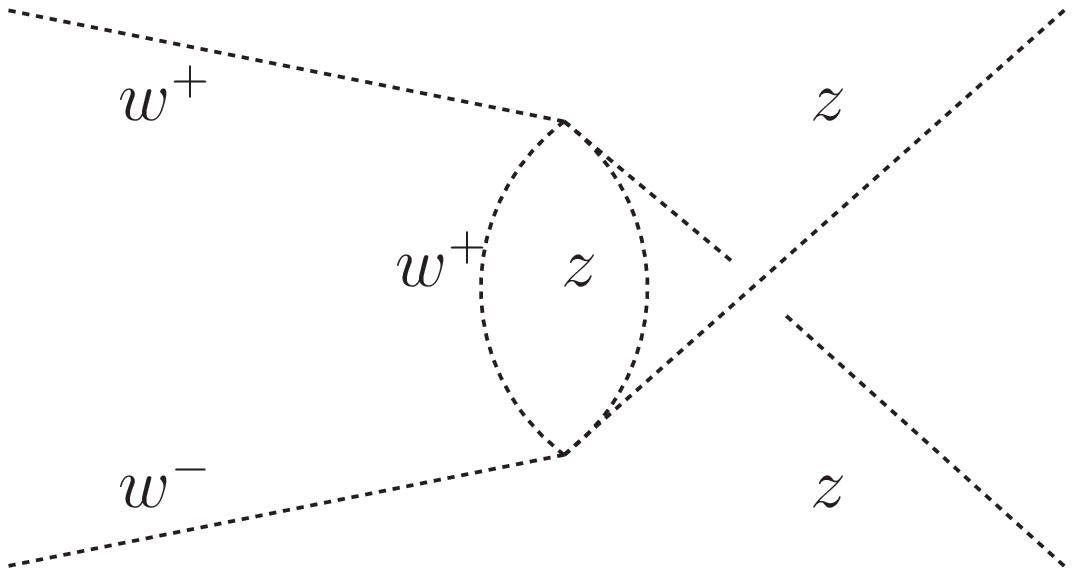}} \\
\caption{The dominant $w^{+} w^{-} \to z z$ Goldstone boson scattering amplitudes at one-loop level in the SM in the limit that $s \gg m_{H}^{2}$.  See Ref.~\cite{DW} for the complete set of one-loop diagrams, valid for all energies, in the limit $s, m_{H}^{2} \gg M_{W}^{2}$}
\label{fig:amplitudes_loop}
\end{figure}

It has been mentioned that a fully analytic expression for the one-loop contribution to $WW$ scattering
is not available. In particular, for the basic building block (related to the other amplitudes 
by isospin and---properly applied---crossing symmetry), there is no closed expression for $W_L^+ W_L^- \to
Z_LZ_L$, i.e. for $A^{+-00}$.
To overcome this difficulty, we will restrict ourselves to the scattering of longitudinally
polarized vector bosons. This is in any case dominant at high energies and it is expected to be the 
most sensitive one to the EWSBS. The restriction to longitudinal $W$ will allow us to determine with
enough precision the loop correction. For our purposes, the loop amplitude for the process $A(W^{+}W^{-} \to Z Z)$ will be calculated in a mixed way, to best approximate the unknown full amplitude. The real part  is determined by direct use of the equivalence theorem~\cite{et,esma}: 
we replace this loop amplitude by the corresponding process $w^+ w^- \to zz$. A proper use of the equivalence 
theorem~\cite{esma} requires keeping the external on-shell condition $(p^a)^2= M_W^2$, as well as in internal lines for consistency. However, the available 
loop calculations that make use of the equivalence theorem all work in the $M_W=0$ limit, which 
makes the calculation much simpler. 
While this is not a major limitation if $s \gg M_W^2$ as will always be the case, terms of the form $(M_W^2/M_H^2)^{2}$ and higher
are potentially missed as well. To have some control on this, we shall use the unitarity relations themselves, which 
become in fact increasingly more and more accurate as $M_H$ is increased. However, unitarity is always very
well satisfied in the results we present below. This gives us confidence in the method used and the 
approximations that we have had to compromise on. 

This calculation can be done in any field parametrization, i.e. it gives the same result using linear 
or nonlinear realizations for the Goldstone boson sectors.  The full result for the amplitude 
is given in Appendix \ref{sec:appendix_GBamplitudes} and is the one
actually used in the numerical result presented in the coming sections. They have
been computed in Refs.~\cite{DW,imaginary}. The result reproduced below is the leading contribution
in the $s \gg M_H^2$ limit, originating from the ``bubble'' diagrams (as seen in Fig.~\cite{fig:amplitudes_loop}) and wave-function renormalization only, and it also agrees with the one found in Ref.~\cite{DW} 
\be
A^{+-00}|_{\rm loop} =  -\frac{\lambda^2}{4\pi^2}\left[ 4 \ln\fracp{s}{M_H^2}+ \ln\fracp{-t}{M_H^2} + \ln\fracp{-u}{M_H^2}
-\frac12 -\frac{9\pi}{2\sqrt{3}}\right] \, ,
\ee
where the Higgs mass is $M_H^2 = 2\lambda v^2$. The imaginary part (of the bubble diagrams only) can be easily guessed
from the above expression.

Several comments are in order. First, because this calculation is done with Goldstone bosons, the above mentioned 
subtleties associated with crossing do not apply. Second this same amplitude in the opposite limit, i.e.
$s \ll M_H^2$, leads to the familiar result
\bea
A^{+-00}|_{\rm loop} & = & 
\fracp{1}{v^2} \fracp{1}{4\pi v}^{2} \Bigg[ \frac{s^2}{2}\ln\fracp{M_H^2}{s}+ \frac{1}{6} t(s+2t)\ln\fracp{M_H^2}{-t} 
\\ \nn & & 
+ \frac{1}{6}u(s+2u) \ln\fracp{M_H^2}{-u}
+ s^{2} \left(\frac{9\pi}{2\sqrt{3}}-\frac{74}{9}\right) -\frac{2}{9}\left(t^2+u^2\right) \Bigg]
\eea
that has been amply used in unitarization analysis for models with a heavy Higgs boson~\cite{heavyhiggs}, with some slight variation in the nonlogarithmic terms due to renormalization-scheme-dependent factors.

As emphasized, the above contribution for the real part is computed by making use of the equivalence theorem in the $M_W=0$ limit, which is approximately valid for large values of $s$.  In addition, the contribution from the effective operators ${\cal L}_i$---parametrized by the coefficients $a_i$---are of order $p^4$ and amply dominate in this limit, making the approximation made in the calculation of the real part of the loop amplitude even less relevant.  We continue to include this contribution, however, to have the best possible control over the amplitude when the $a_{i}$ are taken to be very small.

As for the imaginary part, the situation is very different. When computed with Goldstone bosons using the equivalence theorem as given in Ref.~\cite{imaginary}, it is in fact quite small.  In particular, for the $I=1$ channel, if restriction is kept to the dominant diagrams in the $s \gg M_H^2$ limit (bubbles), one gets zero for the imaginary part. Including the rest of the diagrams that have been computed in Ref.~\cite{imaginary} does not really improve the situation much as they are still much too small.  In order to use unitarity constraints, it is crucial to have good theoretical control on the imaginary parts, and for this reason we have to determine the imaginary parts directly from the tree-level contribution using longitudinal $W$'s rather than Goldstone bosons. 

We can take advantage of one crucial relation to partially circumvent this problem.  We know that the full calculation must satisfy Eq.~(\ref{eq:part_unitarity}) for the fully elastic case and more generally Eq.~(\ref{eq:part_unitarity_hh}) for the inelastic case (appearing only for $I=0$).  We can therefore \textit{define}---without approximation---the imaginary part of $t_{IJ}^{(2)}$ by Eqs.~(\ref{eq:part_unitarity}) and (\ref{eq:part_unitarity_hh}) [and using the isospin amplitudes $T_{I}$ defined in Eqs.~(\ref{eq:fixed_isospin}) and (\ref{eq:fixed_isospin_hh})] without any need for the unknown loop calculation.  Only the real part of the full loop calculation, then, remains approximated, for the lack of a better option, with the real part of the loop amplitude for the scattering of the Goldstone bosons.  However, we believe that for the purpose of identifying dynamical resonances, our calculation should be fairly robust: we know that ${\rm Re \,}t_{IJ}^{(2)}$ must be dominated by the anomalous terms, rather than the loop terms, because in our scenario with the light Higgs boson these alone grow too quickly with energy and are therefore solely responsible for the violation of unitarity.

We now summarize, then, the calculation: in all cases, a fundamental amplitude $A^{+-00}$ is calculated and used to construct the isospin amplitudes $T_{I}$, expressed as the lowest-order partial wave in each channel (i.e. $t_{00}$, $t_{11}$, and $t_{20}$), where

\bea
t_{IJ}^{(0)} & \qquad \to \qquad &  \textrm{calculated from tree-level amplitude with external $W_{L}$} \\  \nn
{\rm Re \,}t_{IJ}^{(2)} & \to & \textrm{calculated from $a_{i}$-dependent terms with external $W_{L}$ + } \\ \nn
& & \textrm{real part of one-loop Goldstone boson scattering amplitudes.} \\ \nn
{\rm Im \,}t_{IJ}^{(2)} & \to & \left\{
\begin{array}{ll}
\sigma(s) |t_{IJ}^{(0)} |^{2} + \sigma_{H}(s) |t_{H,IJ}^{(0)}|^{2} \hspace{1cm} & \textrm{if $I=0$} \\ 
\sigma(s) |t_{IJ}^{(0)} |^{2} & \textrm{otherwise} \\
\end{array}
\right. \\ \nn
\eea

\noindent The final partial wave amplitudes, $t_{IJ}$, when defined by the IAM according to Eq.~(\ref{eq:t_iam}), will necessarily satisfy the perturbative unitarity constraints to the order with which we are working \textit{by construction} and have been explicitly verified in our numerical results.

%%%%%%%%%%%%%%%%%%%%%%%%%%%%
\section{Resonances} \label{sec:resonances}
%%%%%%%%%%%%%%%%%%%%%%%%%%%%
As in the earlier calculations with Higgs-less models, we can identify dynamical resonances appearing in our unitarized amplitudes by searching for places where the phase shifts of the amplitudes, $\delta_{IJ}$, pass through $(\pi / 2)$ or, equivalently, when $\cot \delta_{IJ}$ passes through zero with a negative slope.  We must, however, forbid any region of parameter space in which any amplitude develops a ``resonance'' that has a phase shift crossing $- (\pi /2)$, which would imply an unphysical, negative decay width.  We will call these ``false resonances.''

\begin{figure}[tb]
\centering
\subfigure[]{\includegraphics[clip,width=0.45\textwidth]{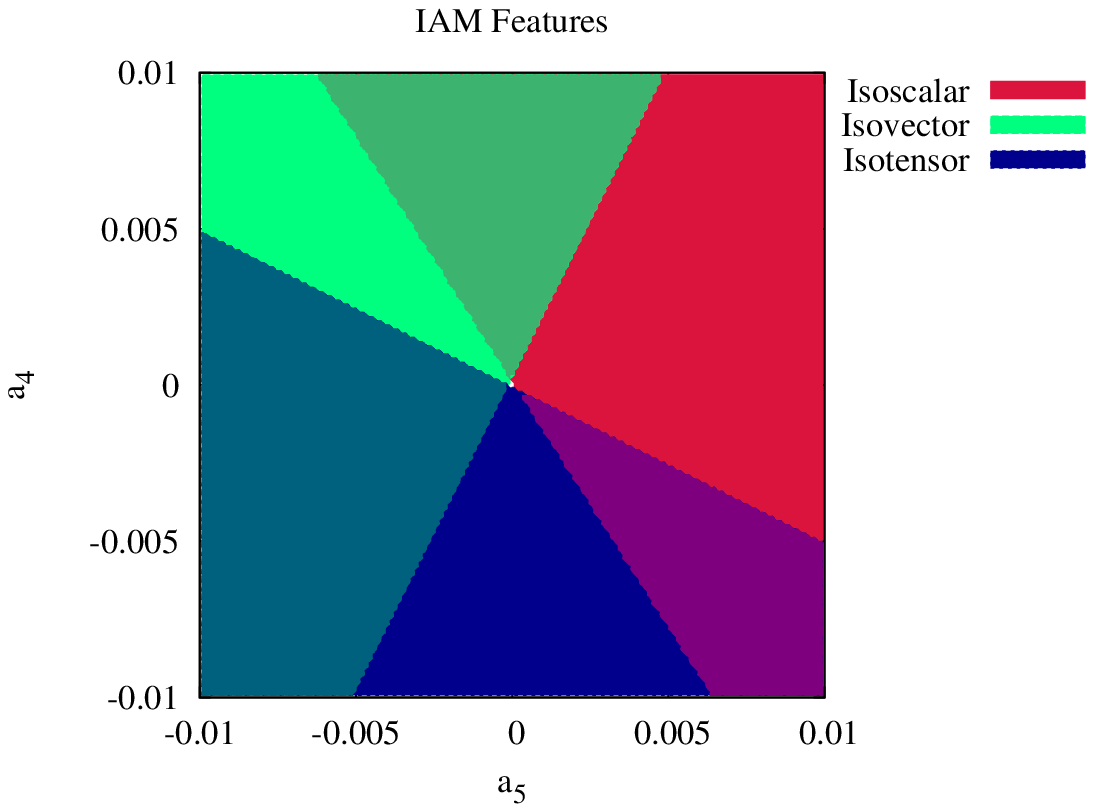}}
\subfigure[]{\includegraphics[clip,width=0.45\textwidth]{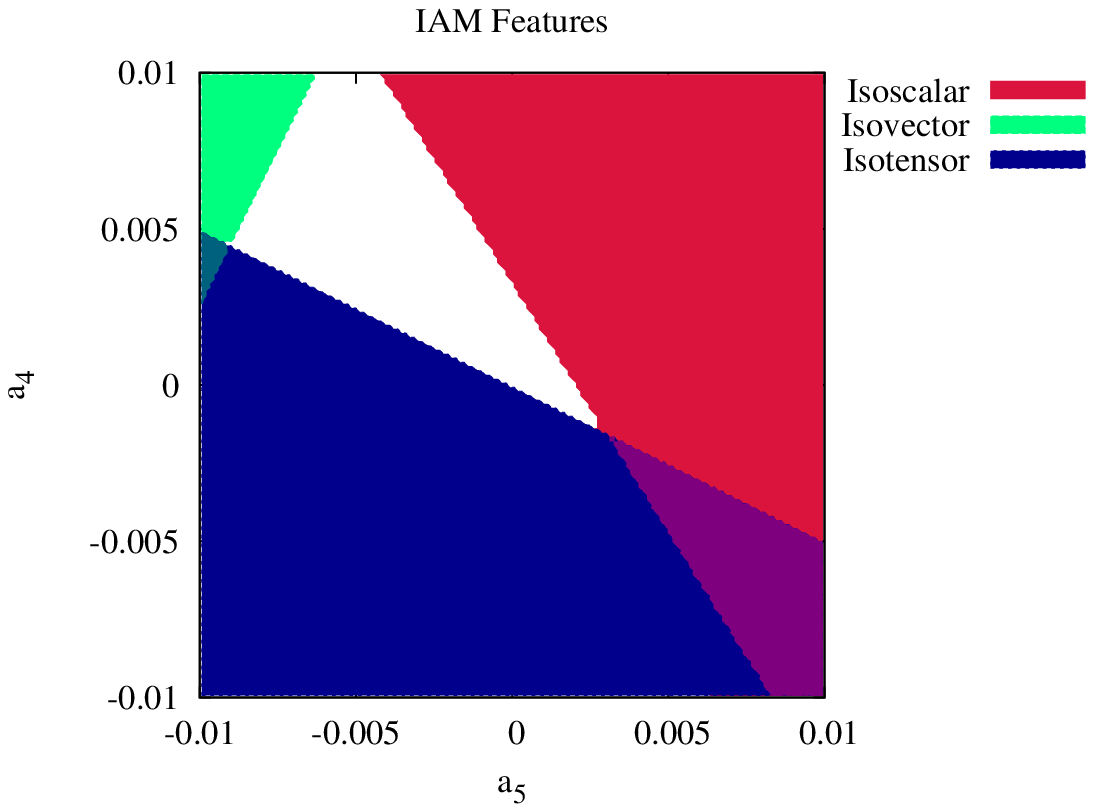}} \\
\caption{(a) Regions with isoscalar and isovector resonances (and the isotensor \textit{exclusion} region) up to a scale $4\pi v \approx 3$~TeV.  (b) Same as (a), but only showing isoscalar/isovector resonances in which $M_{S,V}<600$~GeV, for comparison with Higgs search results.}
\label{fig:resonances}
\end{figure}

In Fig.~\ref{fig:resonances}(a), we present the results for our search in the $a_{4}-a_{5}$ parameter space for $-0.01 < a_{4},a_{5} < 0.01$.  We have imposed the usual cutoff in the search of $\sqrt{s} = 4\pi v\simeq 3$~TeV.  We find, as in previous work, that there is a region (shown in red) where there are only scalar resonances, a region (in green) where there are only vector resonances, an overlapping region where there are both, and finally a large region (in blue) in which the isotensor amplitude develops unphysical, false resonances and therefore must be excluded.  There is also a small region, centered around $a_{i}=0$, in which there are no resonances or unphysical features to be found.  Such a region was also found in the earlier, Higgs-less work, though it was notably larger \cite{technicolor,butterworth}.  Its interpretation here, however, is quite different: contrary to previous work, values of $a_{i}=0$ correspond here to the SM with a light Higgs boson, a theory that suffers no problems with unitarity and therefore should not be expected to develop dynamical resonances from this method.  The absence of any features is a good check that the IAM is not introducing them when it should not.

We can understand the results of Fig.~\ref{fig:resonances}(a) as the following: Typically an extended symmetry breaking scenario has more resonances than just a light ''Higgs'' boson. There could be additional scalars (such as the ones appearing for instance in an $SO(6)/SO(5)$ model~\cite{composite}), vector resonances, or even higher spin states. The low-energy contribution from these states is parametrized by the $a_{i}$. Figure~\ref{fig:resonances}(a), then, addresses the following question: What do we exclude if we assume that {\it no} additional resonance is seen anywhere between the state at 125 GeV and $4\pi v\simeq 3$~TeV? The excluded region, then, in $a_{4} - a_{5}$ parameter space looks very dramatic. Only values extremely close to zero are acceptable, reflecting of course that the new states must be quite heavy and perhaps beyond the reach of our method. 

Let us now examine which are the {\it current} bounds, i.e. the exclusion region for $a_{4}$ and $a_{5}$ that can be obtained by assuming that no new resonances exist below 600 GeV (but that may yet exist above the currently unexplored regions), as probed by the published Higgs search data for $\wplus \wminus$ and $ZZ$ decay modes. This is shown in Fig.~\ref{fig:resonances}(b), where this limit is placed only on the physical resonances of the isoscalar/isovector channels.  These exclusion regions assume, however, that these resonances would have signals with strengths comparable to that of a SM Higgs boson of the same mass.  The viability of this assumption will be addressed in the next section.

\begin{figure}[tb]
\centering
\subfigure[]{\includegraphics[clip,width=0.45\textwidth]{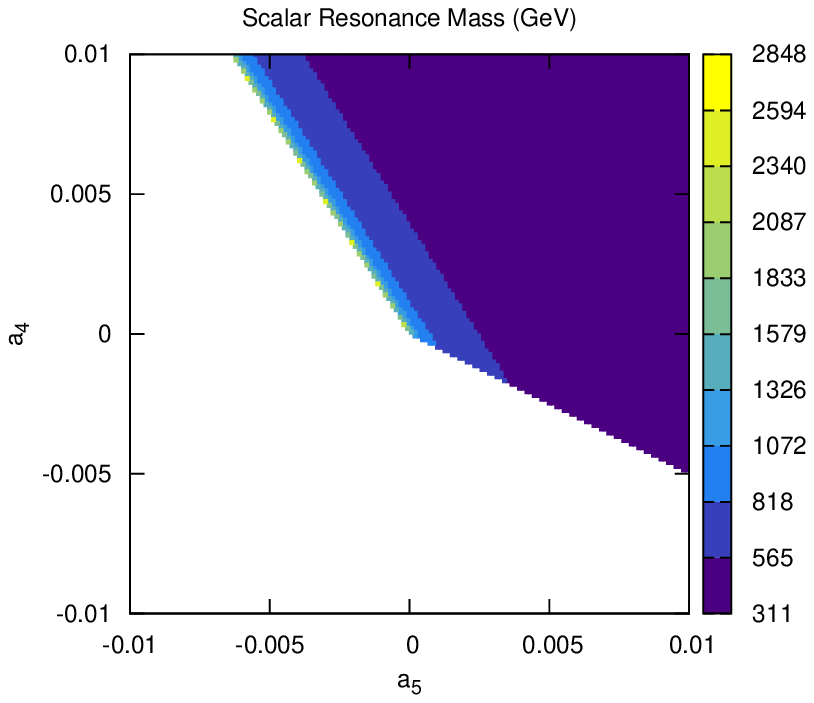}} \hspace{0.5cm}
\subfigure[]{\includegraphics[clip,width=0.45\textwidth]{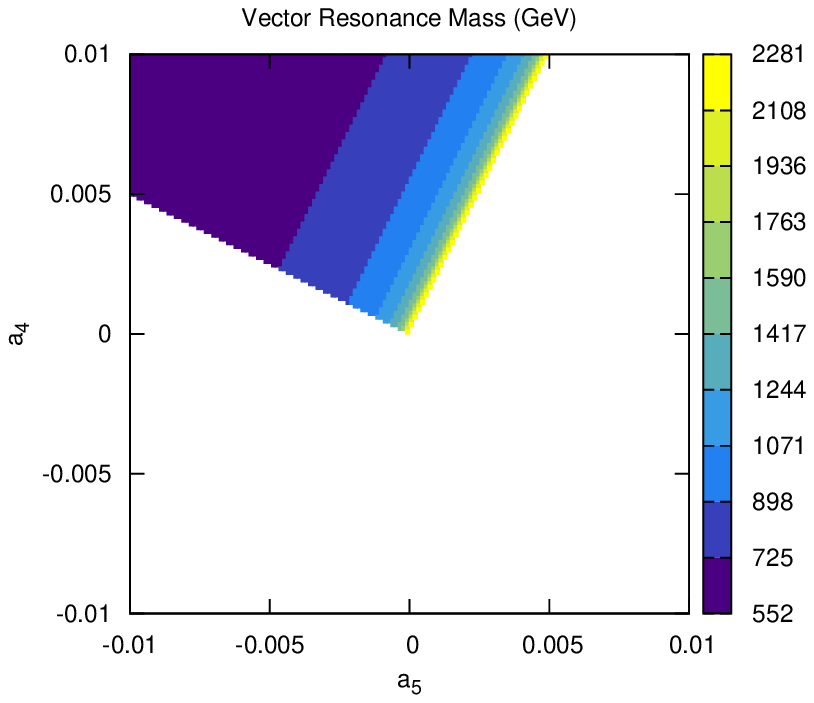}} \\
\subfigure[]{\includegraphics[clip,width=0.45\textwidth]{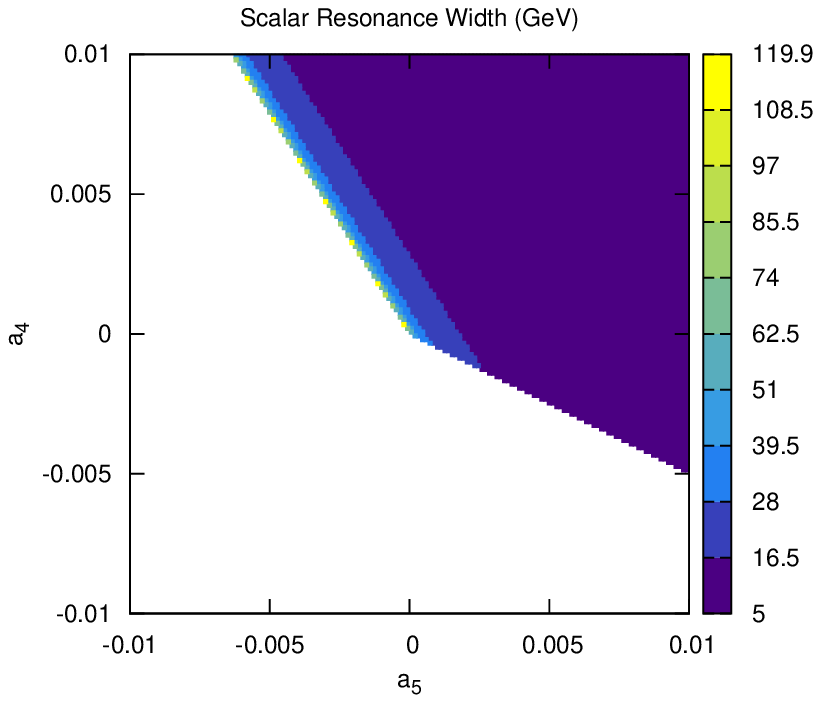}} \hspace{0.5cm}
\subfigure[]{\includegraphics[clip,width=0.45\textwidth]{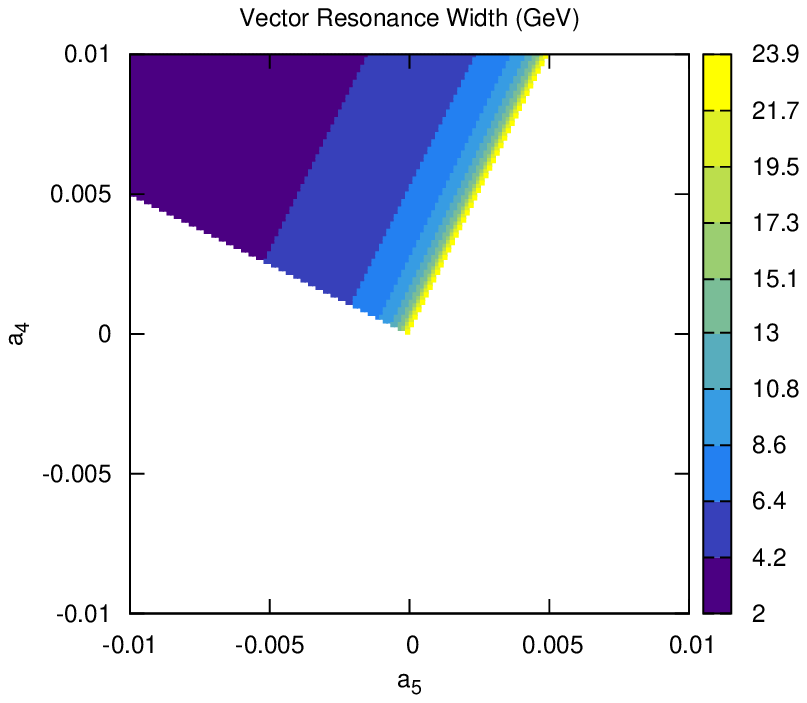}} \\
\caption{Masses in GeV for (a) scalar and (b) vector resonances predicted from the unitarized partial wave amplitudes of $WW \to WW$ scattering.  Widths in GeV for the corresponding (c) scalar and (b) vector resonances.}
\label{fig:masses}
\end{figure}

In Figs.~\ref{fig:masses}(a)-(d), we give contours for the predicted masses and widths of the isoscalar/isovector resonances over the $a_{4}-a_{5}$ parameter space.  To estimate the widths, we continue our amplitudes into their second Riemann sheet and solve for the complex pole such that 
\be
t_{IJ}^{-1}(s_{pole}) = 0 \, ,
\ee
where $s_{pole}$ is interpreted as 
\be
s_{pole} = \left( m_{pole}^{2} - i \, m_{pole} \Gamma_{pole} \right) \, .
\ee

These pole masses and widths are what are plotted in Figs.~\ref{fig:masses}~(a)-(d), and we note that for large estimated widths, these pole masses may diverge slightly from those predicted by the location where the phase shifts $\delta_{IJ}$ pass through $(\pi / 2)$.  What we can see in these figures is that the predicted scalar masses range from $\sim 300$~GeV to nearly the cutoff of $3$~TeV and the vectors from $\sim 550$~GeV to $\sim 2.3$~TeV, generally lower than what would be predicted in the Higgs-less theories.  The widths are particularly interesting: except for the largest masses, they are $\mathcal{O}(1$~GeV) to $\mathcal{O}(10$~GeV).  This is noticeably more narrow than the widths predicted in the Higgs-less theories, which are typically $\mathcal{O}(100$~GeV) over much of the parameter space.  The inclusion of a light Higgs-like state, then, clearly alters the characteristics of the resonances produced by the IAM.

%%%%%%%%%%%%%%%%%%%%%%%%%%%%
\section{Cross Sections} \label{sec:xsec}
%%%%%%%%%%%%%%%%%%%%%%%%%%%% 
To address the issue of just how strong the signals coming from the dynamical resonances would be, we need to calculate observable cross sections for the longitudinal vector boson scattering amplitudes (e.g. $A^{+-00}$), converting back from the now unitarized partial waves $t_{IJ}$.  We can do so employing the inverse procedure of Sec.~\ref{sec:isospin}, following closely that of Ref.~\cite{butterworth}.  The fixed-isospin amplitudes are formally defined in terms of the $t_{IJ}$ as
\be
T_{I} = 32\pi \sum_{J=0}^{\infty} (2J+1) t_{IJ} P_{J} (\cos\theta) \, .
\ee
We now ignore the higher partial waves and use only the lowest-order amplitude for each 
isospin channel of interest, such that
\bea
T_{0} & \approx & 32\pi t_{00} \\ \nn
T_{1} & \approx & 32\pi (3 t_{11} \cos\theta) \\ \nn
T_{2} & \approx & 32\pi t_{20} \, . 
\eea
We will be concerned here with the observable amplitudes $A(W^{+}W^{-} \to W^{+}W^{-})$ and 
$A(Z Z \to Z Z)$, which can be defined in terms of the $T_{I}$ as
\bea
A(W^{+}W^{-} \to W^{+}W^{-}) &=& \frac{1}{3} T_{0} + \frac{1}{2} T_{1} + \frac{1}{6} T_{2} \\ \nn
A(Z Z \to Z Z)            &=& \frac{1}{3} T_{0}                    + \frac{2}{3} T_{2} \,  .
\eea
Finally, we must relate these amplitudes to the detectable LHC cross sections $\sigma(pp \to WW jj)$, and 
in doing so we will employ the EWA (effective W approximation) \cite{ewa} even if we are aware
that it is applicable at much larger energies only.\footnote{The unitarization of $W_{L} W_{L} \to W_{L} W_{L}$ in the process $pp \to WW jj$ has indeed been analyzed in full Monte Carlo simulations which include not only the full 6 fermion final states but the interference with the amplitudes for the transversely polarized states as well.  These works find that the resonances generated with the IAM and similar methods are still detectable, albeit less pronounced, and depend heavily on rapidity cuts on the tagged jets.  Furthermore, significant discrepancies in the invariant mass distribution may be found outside the peak region \cite{sixfermion}.}  However, we are only after a guidance of the
relative strength of the different signals, and the EWA---which is technically simple to use---should
most likely suffice. For the $W W$ scattering amplitude defined in the $W W$ rest frame as
\be
\frac{d \sigma_{WW}}{d \cos\theta} = \frac{|A|^{2}}{32\pi M^{2}_{WW}} \, ,
\ee
the corresponding LHC cross section is given by 
\be
\frac{d \sigma}{d M^{2}_{WW}} = \sum_{i,j} \int_{M^{2}_{WW}}^{1} \int_{M^{2}_{WW}/(x_{1} s)}^{1} \frac{dx_{1} dx_{2}}{x_{1}x_{2}s}
f_{i}(x_{1},\mu_{F}) f_{j}(x_{2},\mu_{F}) \frac{dL_{WW}}{d\tau} \int_{-1}^{1} \frac{d \sigma_{WW}}{d \cos\theta} d\cos\theta \, ,
\ee
where $\tau = \hat{s}/s = M^{2}_{WW}/(x_{1}x_{2} s)$ and where $\sqrt{s}=8$~TeV for our current analysis.  
We set the factorization scale, $\mu_{F}$, to the $W$-boson mass and use the CTEQ6L1 parton distribution functions \cite{pdf}.
The effective luminosity for longitudinal $W$ and $Z$ bosons is given as
\be
\frac{d L_{WW}}{d\tau} = \fracp{g}{4\pi}^{4} \fracp{1}{\tau}\left[ 
(1+\tau) \ln\fracp{1}{\tau} - 2 (1-\tau) 
\right] \, .
\ee
A factor of $(1/2)$ should also be included in the final expression for the $ZZ \to ZZ$ amplitude to account for the identical particle in the final state.

\begin{figure}[tb]
\centering
\subfigure[]{\includegraphics[clip,width=0.45\textwidth]{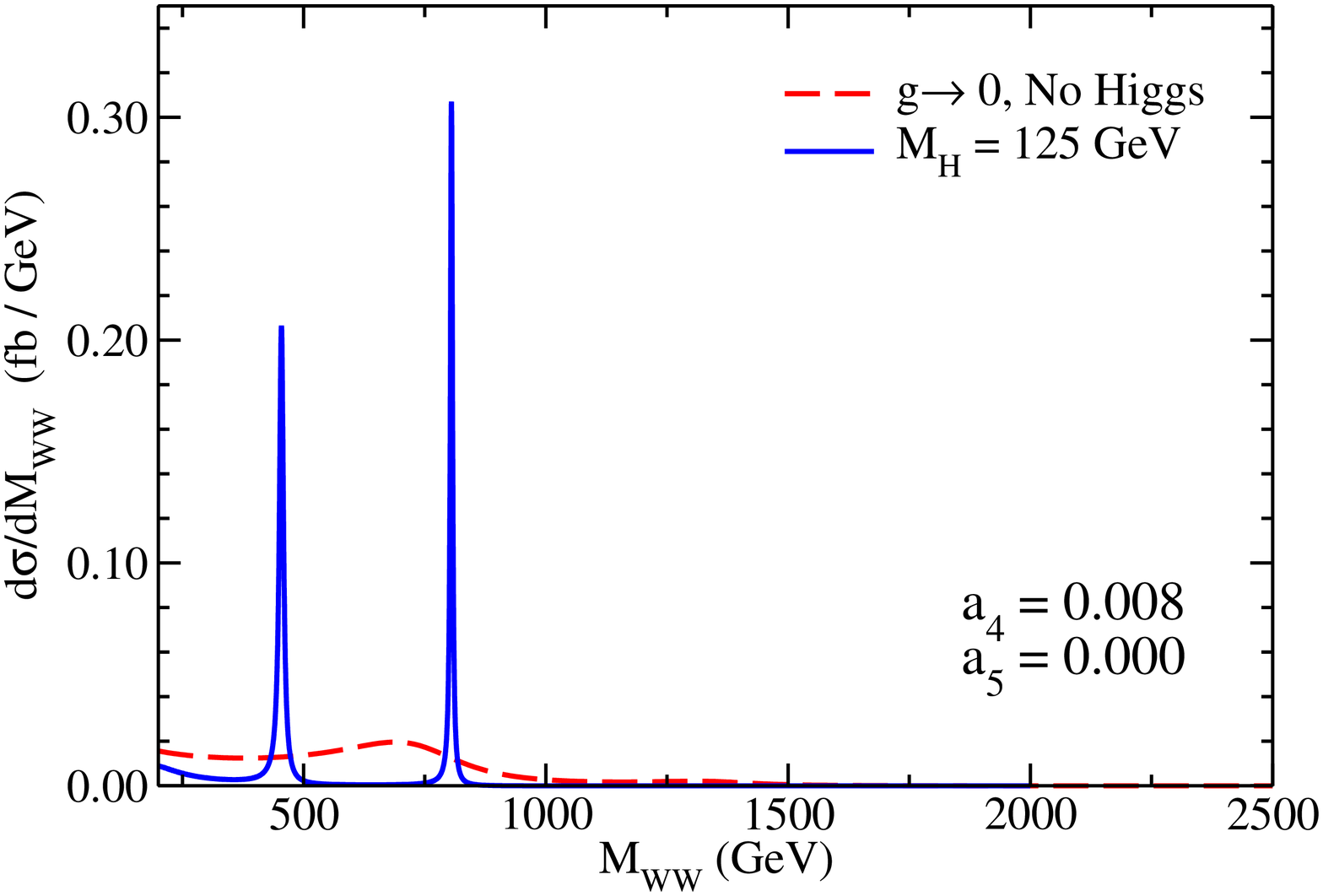}} \hspace{0.5cm}
\subfigure[]{\includegraphics[clip,width=0.45\textwidth]{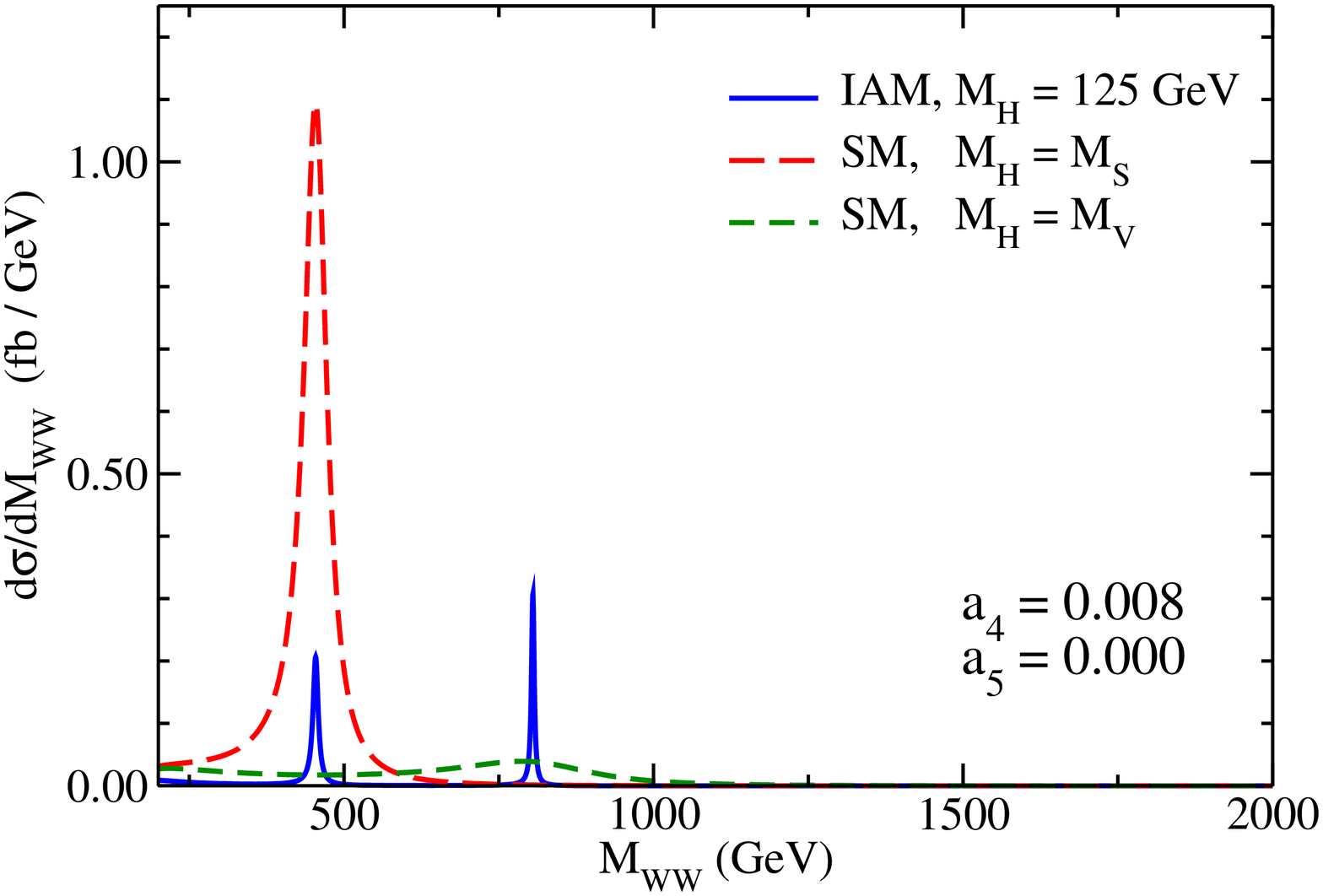}} \\
\caption{(a) The red, dashed curve gives LHC cross section results at $\sqrt{s} = 8$~TeV resulting from the IAM for the Higgs-less case with $g \to 0$, $a_{4}(1~{\rm TeV})=0.008$, and $a_{5}(1~{\rm TeV})=0.000$ (Point D from Ref.~\cite{butterworth}).  The solid blue curve gives the updated results for finite $g$ and $M_{W}$ and a Higgs mass of $M_{H}=125$~GeV. (b) The red, long-dashed curve gives the tree-level SM result for a Higgs mass equal to that of the scalar resonance ($\sim 454$~GeV), the green, short-dashed curve gives the corresponding result for a Higgs mass equal to the vector resonance ($\sim 805$~GeV), and the solid, blue curve is the same as in (a).}
\label{fig:xsec}
\end{figure}

With these expressions, we can now estimate the signal strength of the $WW$ resonances described in Sec.~\ref{sec:resonances}.  We give one explicit example in Fig.~\ref{fig:xsec}(a), where we plot the differential cross section for $\wplus \wminus \to \wplus \wminus$ as a function of $M_{WW}$ for the coefficient values $a_{4}=0.008$ and $a_{5}=0.000$.  This corresponds to benchmark Point D in Ref.~\cite{butterworth} (taken at scale $\mu=1$~TeV), chosen here to demonstrate generic features in a region of parameter space where both scalar and vector resonances are present.  We also plot the earlier results of that paper---calculated with the Goldstone amplitudes and no light Higgs boson---in Fig.~\ref{fig:xsec}(a) for comparison.  A primary observation is that while both the earlier and updated results predict both a scalar and vector resonance for this choice of parameters, the new resonances appear with lighter masses and significantly smaller widths, as discussed in the previous section.  These, in turn, may translate into LHC signals that are easier to detect than those considered in studies such as Refs.~\cite{technicolor,butterworth}.

The question, then, arises: if any of these resonances exist, should they have already been seen in the Higgs search data in the $\wplus \wminus$ and $ZZ$ decay channels, along with the ``Higgs boson'' itself?  To answer this, we will construct approximate comparisons of the relevant signal strengths.  The Higgs search data as published in Refs.~\cite{atlas,cms} are summarized by exclusion limits in $(\sigma / \sigma_{SM})$ at a given mass $M_{H}$, where $\sigma$ is the observed signal strength and $\sigma_{SM}$ is the SM prediction for a Higgs boson of mass $M_{H}$ and corresponding SM decay width.

For our purposes, we will define the cross section coming from the resonance region for a given resonance of mass $M_{R}$ and width $\Gamma_{R}$ as
\be
\label{eq:peak_r}
\sigma^{peak}_{R} \equiv \int_{M_{R} - 2 \Gamma_{R}}^{M_{R} + 2 \Gamma_{R}} \left[d M_{WW} \times \frac{d\sigma_{R}}{d M_{WW}} \right] \, ,
\ee
where $\sigma_{R}$ is the cross section resulting from the amplitudes unitarized by the IAM.  
For a SM Higgs boson with mass $M_{H}$ set to $M_{R}$ and corresponding SM decay width $\Gamma_{H}$, we also calculate
\be
\label{eq:peak_sm}
\sigma^{peak}_{SM} \equiv \int_{M_{H} - 2 \Gamma_{H}}^{M_{H} + 2 \Gamma_{H}} \left[ d M_{WW} \times \frac{d\sigma_{SM}}{d M_{WW}} \right] \, ,
\ee
where here $\sigma_{SM}$ is calculated at tree level with the appropriate Higgs mass, whose width 
is included via the replacement $M_{H}^{2} \to \left(M_{H}^{2} - i M_{H} \Gamma_{H} \right)$.  Using this information, we then define the ratio
\be
R^{peak} \equiv \fracp{\sigma^{peak}_{R}}{\sigma^{peak}_{SM}},
\ee
which is a function of the coefficients $a_{i}$.  In addition to providing a variable to compare with the Higgs search data, this quantity has the benefit of potentially mitigating any problematic effects resulting from the use of the EWA and from only considering the contributions from the scattering of the longitudinal components of the vector bosons.

Before showing the results for $R^{peak}$ over the $a_{4} - a_{5}$ parameter space, we first demonstrate in Fig.~\ref{fig:xsec}(b) the type of comparison we are making.  Here, the results from the IAM are presented once again at benchmark Point D, and we now compare them with the SM calculation using Higgs boson masses set to those of the scalar and vector resonances.  At lighter masses (in this case, that of the scalar), a corresponding Higgs signal would still be much more visible than that of these new dynamical resonances.  It should be noted, however, that at higher masses, such as that of the vector resonance in this figure, the Higgs width becomes very broad, making the direct comparison less obvious as its signal is more diluted.

\begin{figure}[tb]
\centering
\subfigure[]{\includegraphics[clip,width=0.45\textwidth]{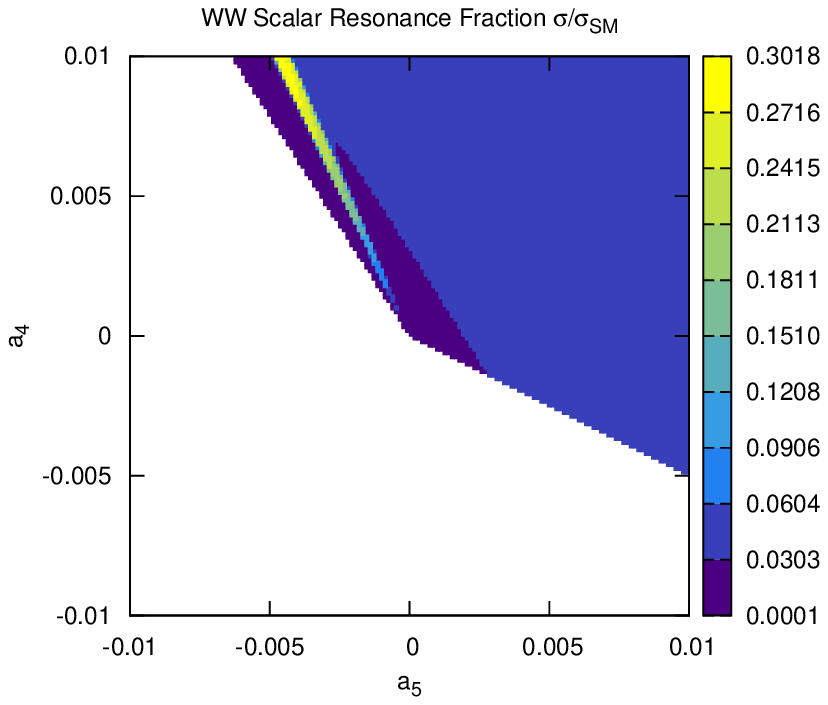}} \hspace{0.5cm}
\subfigure[]{\includegraphics[clip,width=0.45\textwidth]{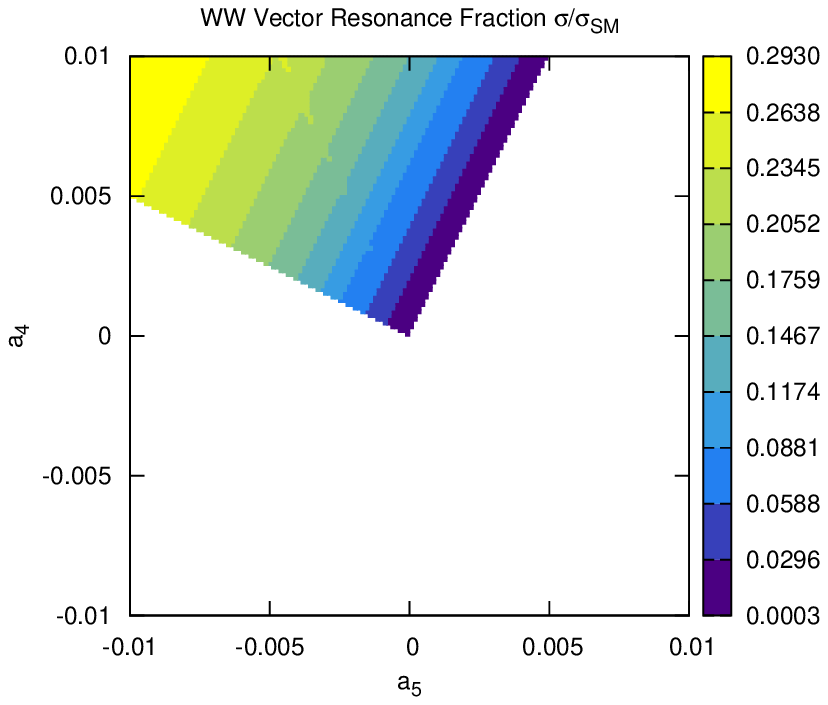}} \\
\subfigure[]{\includegraphics[clip,width=0.45\textwidth]{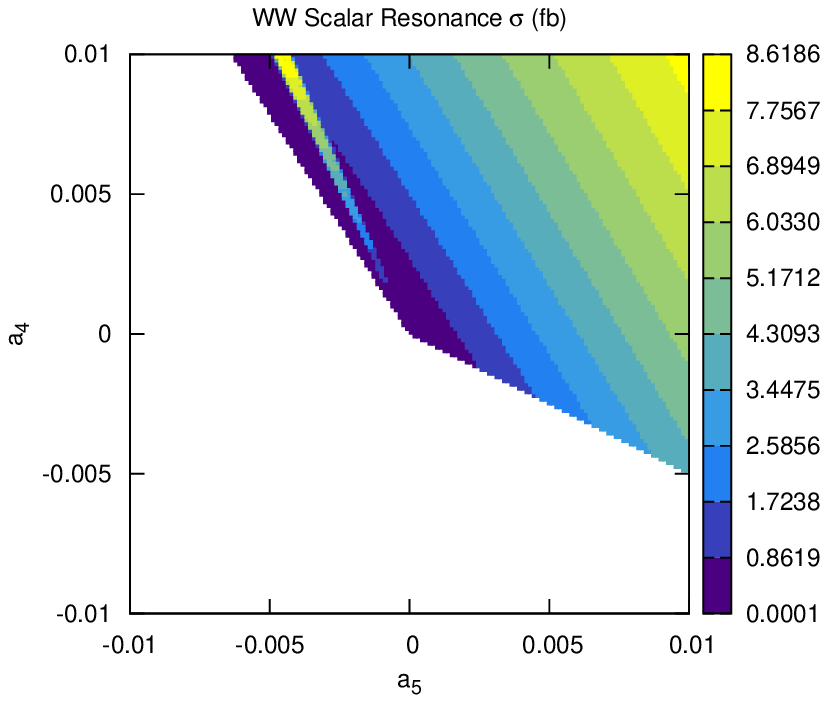}} \hspace{0.5cm}
\subfigure[]{\includegraphics[clip,width=0.45\textwidth]{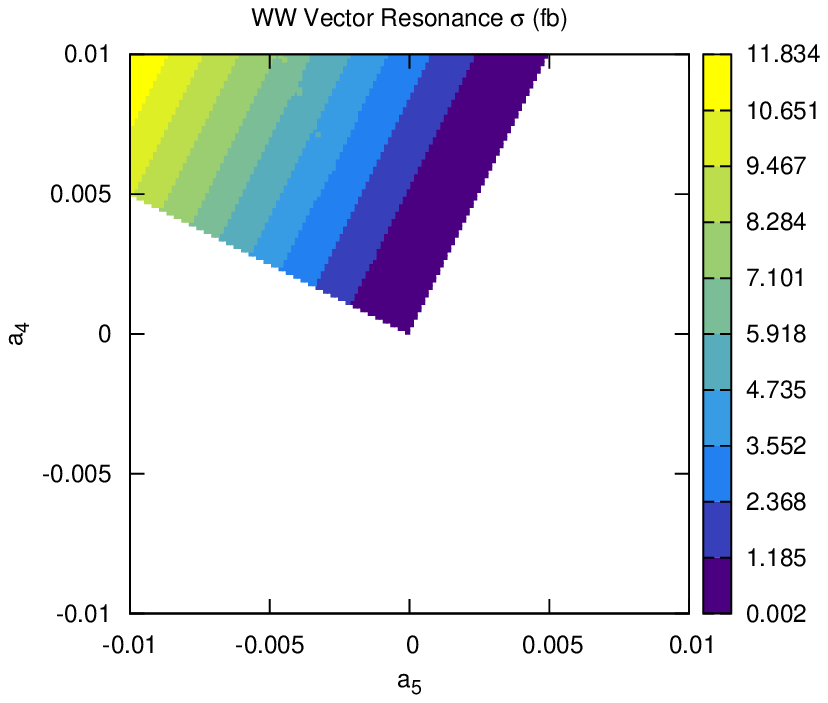}} \\
\caption{Ratio of $WW$ scattering cross section due to dynamical resonances with that of the SM with a Higgs boson of the same mass for (a) scalar and (b) vector resonances, taken in the peak region as defined in the text.  The resonance cross sections are given in fb for the isoscalar resonances in (c) and isovector resonances in (d).}
\label{fig:signal_ww}
\end{figure}

\begin{figure}[tb]
\centering
\subfigure[]{\includegraphics[clip,width=0.45\textwidth]{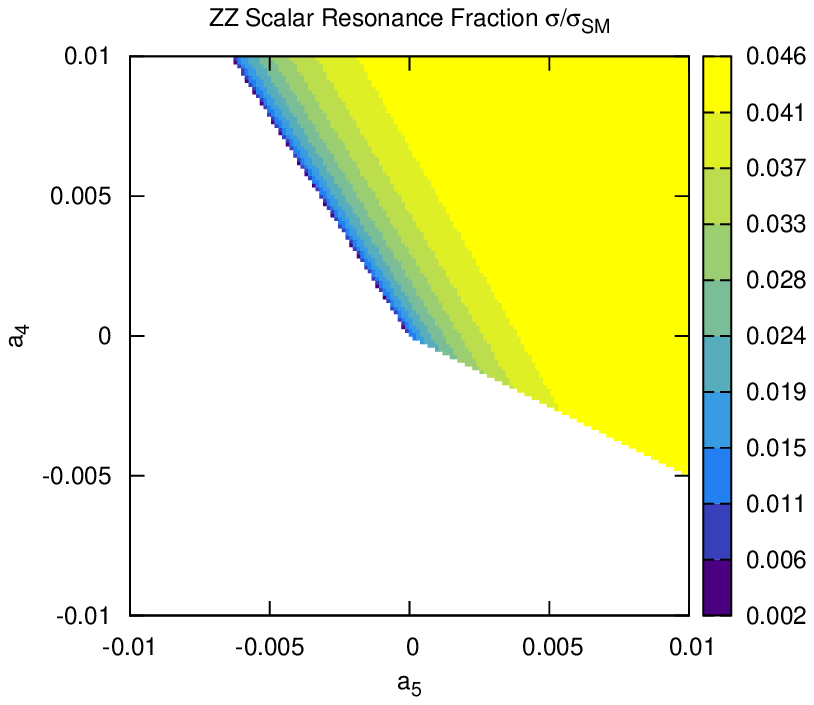}} \hspace{0.5cm}
\subfigure[]{\includegraphics[clip,width=0.45\textwidth]{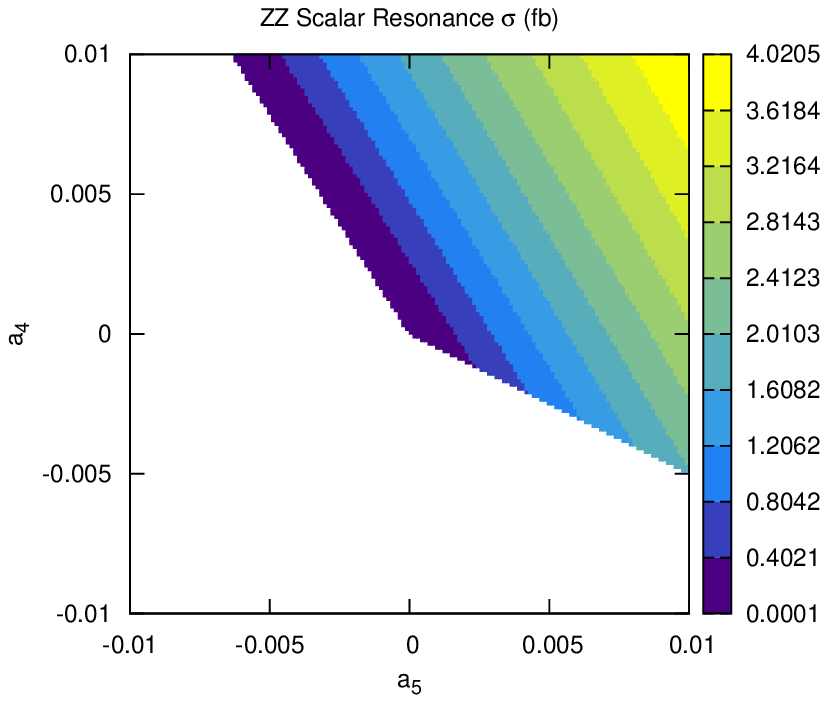}} \\
\caption{(a) Ratio of $ZZ$ scattering cross section due to dynamical scalar resonances with that of the SM with a Higgs boson of the same mass, taken in the peak region as defined in the text.  The resonance cross sections are given in fb in (b).
}
\label{fig:signal_zz}
\end{figure}

The values for the relative signal strengths in the $\wplus \wminus \to \wplus \wminus$ channel for the scalar and vector resonances are given in Figs.~\ref{fig:signal_ww}(a) and \ref{fig:signal_ww}(b), respectively.  For completeness, we also include the values for $\sigma^{peak}$ in units of femtobarns in Figs.~\ref{fig:signal_ww}(c) and \ref{fig:signal_ww}(d).  It is notable that over all the parameter space, the results are $\mathcal{O}(0.1)$ or lower, with a maximum strength of $\sim 0.3$.  It is clear from the Higgs search data, however, that the values for $(\sigma / \sigma_{SM})$ that are currently being probed are, at best, $\mathcal{O}(1)$, particularly when only the vector boson fusion channels are extracted from the data.  This suggests that resonances such as these would \textit{not}, in fact, be currently probed by the existing Higgs data in the $\wplus \wminus \to \wplus \wminus$ channel.

One interesting feature should be noted in these data, however.  There is an obvious strip of higher cross sections visible in both the scalar and vector results.  In this region, the scalar and vector resonances are in fact \textit{mass degenerate}, or at least approximately so.  The results are then essentially the sum of the individual channel results.  A close comparison of Figs.~\ref{fig:masses}(a) and \ref{fig:masses}(b) also shows this result and that, furthermore, there is a small region in which the scalar is heavier than the vector resonance, corresponding to the region to the left of the enhanced strip in Fig.~\ref{fig:signal_ww}.

In Figs.~\ref{fig:signal_zz}(a) and \ref{fig:signal_zz}(b), we also give the results for $R^{peak}$ and $\sigma_{R}$ for the possible scalar resonances in the $ZZ \to ZZ$  channel (as there are no isovector resonances possible in this channel).  They are similar to those of the $\wplus \wminus \to \wplus \wminus$ channel, but lack the overlap region due to the lack of vector resonances.  They suggest, however, that the current LHC data is also insensitive to these dynamical resonances in this channel, should they exist.  While a full Monte Carlo simulation would be necessary to accurately estimate the amount of data needed to exclude the existence of these states, in either channel, suffice it to say that after the next shutdown and upgrade, the LHC experiments will eventually acquire the (roughly an order of magnitude more) data needed to almost fully probe the parameter space leading to these states.

%%%%%%%%%%%%%%%%%%%%%%%%%%%%
\section{Conclusion} \label{sec:conclusion}
%%%%%%%%%%%%%%%%%%%%%%%%%%%% 

In light of the discovery of what appears to be a very SM-like Higgs boson, we have revisited the scattering of longitudinal vector bosons at 
high energies and the notion of the apparent violation of perturbative unitarity.  While the SM itself does not suffer from these problems 
like earlier Higgs-less theories, the introduction of small deviations in the vector boson couplings that would result from the 
low-energy contributions of a more complicated, higher-energy EWSBS sector can reintroduce them.  

We have calculated the scattering amplitudes using the longitudinal components of the vector bosons themselves, 
rather than the corresponding Goldstone bosons, and enforced perturbative unitarity through the use of the inverse amplitude method. 
We have performed a number of checks indicating that our methods are valid in the region where applied. The amplitudes are unitary
by construction and reproduce very well the perturbative expansion even at relatively low energies.
 
We have found that, even when including a light SM Higgs boson of mass $M_{H} = 125$~GeV, the present analysis predicts the appearance 
of dynamical resonances in much of the parameter space of the higher-order coefficients.  The masses of these resonances 
extend from as low as $300$~GeV to nearly as high as the cutoff of the method of $3$~TeV, with rather narrow widths typically of 
order 1--10~GeV.  In the absence of observing these resonances up to the cutoff of $\sim 3$~TeV, nearly the entire parameter space 
of the anomalous couplings could be excluded.  We show, however, that the actual signal strength of these resonances, when 
compared with current Higgs search data, indicates that they are not currently being probed in LHC Higgs search data, typically 
giving signals an order of magnitude or 2 weaker than would a SM Higgs boson of the same mass.  Nevertheless, if these 
anomalous vector boson couplings exist, the resulting dynamical resonances they predict should be observable with future LHC data.

We find it interesting that it is the unitarization of the scattering amplitudes that a light Higgs boson brings about that changes profoundly the
resonance structure with respect to the Higgs-less (or a very heavy Higgs) scenario in extended scenarios of EWSBS.  We note, in particular, the resulting reduction in the masses and widths of the resonances that would typically be predicted in the earlier Higg-less theories.

In this work the conservative assumption of taking the Higgs couplings to the light sector to be exactly the same ones as in the minimal SM has been adopted. It would of course be very interesting to relax this hypothesis and consider slightly more general cases and see what effect the resulting tree-level unitarity violation would have on the results presented here.  An additional extension to this work could include performing a complete Monte Carlo simulation of the $pp \to WW jj $ amplitude, without recourse to the effective $W$ approximation, to determine more accurately the observability of these resonance signals at the LHC.

%%%%%%%%%%%%%%%%%%%%%%%%%%%%
\section*{Acknowledgments}
%%%%%%%%%%%%%%%%%%%%%%%%%%%% 
Several conversations with J.Bernabeu and O.Vives triggered our interest in revisiting the unitarization of $W_L W_L$ scattering. We are particularly indebted to J.R.Pel\'aez for having shared his insight on the IAM method with us.  We gratefully acknowledge the financial support from MICINN project FPA2010-20807, DEC project 2009SGR502, and CPAN (Consolider CSD2007-00042).

\newpage
%%%%%%%%%%%%%%%
\bibliographystyle{h-physrev}
%\newpage
\bibliography{ms.bbl}

\begin{thebibliography}{99}


\bibitem{atlas} G.~Aad {\it et al.} (ATLAS Collaboration), Phys.\ Lett.\ B {\bf 716}, 1 (2012). 

\bibitem{cms} S.~Chatrchyan {\it et al.} (CMS Collaboration), Phys.\ Lett.\ B {\bf 716}, 30 (2012).

\bibitem{spin} M.~Zanetti (on behalf of the CMS Collaboration), talk at HCP2012, Kyoto.

\bibitem{yang} L.~D.~Landau, Dokl.\ Akad.\ Nauk Ser.\ Fiz.\  {\bf 60}, 207 (1948); C.~N.~Yang, Phys.\ Rev.\ {\bf 77}, 242 (1950).

\bibitem{twophotons} G.~Aad {\it et al.} (ATLAS Collaboration), Report No.~ATLAS-CONF-2012-168; S.~Chatrchyan {\it et al.} (CMS Collaboration), Report No.~CMS-HIG-12-015 .

\bibitem{exclusion} S.~Chatrchyan {\it et al.}  (CMS Collaboration), Phys.\ Lett.\ B {\bf 710}, 26 (2012); G.~Aad {\it et al.}  (ATLAS Collaboration), Phys.\ Lett.\ B {\bf 716}, 1 (2012); K. Einsweiler (on behalf of the ATLAS collaboration), talk at the HCP Symposium, Kyoto, 2012; C. Paus (on behalf of the CMS collaboration),  talk at the HCP Symposium, Kyoto, 2012.

\bibitem{twodoublets} D.~Toussaint, Phys.\ Rev.\ D {\bf 18}, 1626 (1978); H.~Georgi, Hadronic J.\  {\bf 1}, 155 (1978).

\bibitem{ec} P.~Ciafaloni and D.~Espriu, Phys.\ Rev.\ D {\bf 56}, 1752 (1997)
 
\bibitem{composite} K.~Agashe, R.~Contino and A.~Pomarol, Nucl.\ Phys.\ B {\bf 719}, 165 (2005); R.~Contino, L.~Da Rold and A.~Pomarol, Phys.\ Rev.\ D {\bf 75}, 055014 (2007); B.~Gripaios, A.~Pomarol, F.~Riva and J.~Serra, J.~High Energy Phys.~04 ({\bf 2009}) 070.

\bibitem{couplings} For a recent update see T.~Corbett, O.~J.~P.~Eboli, J.~Gonzalez-Fraile and M.~C.~Gonzalez-Garcia, Phys.\ Rev.\ D {\bf 87}, 015022 (2013), and references therein.

\bibitem{iam} T.~N.~Truong, Phys.\ Rev.\ Lett.\  {\bf 61}, 2526 (1988); A.~Dobado, M.~J.~Herrero and T.~N.~Truong, Phys.\ Lett.\ B {\bf 235}, 134 (1990); A.~Dobado and J.~R.~Pel\'aez, Phys.\ Rev.\ D {\bf 47}, 4883 (1993); J.~A.~Oller, E.~Oset and J.~R.~Pel\'aez, Phys.\ Rev.\ Lett.\  {\bf 80}, 3452 (1998); Phys.\ Rev.\ D {\bf 59}, 074001 (1999); {\bf 60}, 099906(E) (1999); {\bf 75}, 099903(E) (2007); F.~Guerrero and J.~A.~Oller, Nucl.\ Phys.\ B {\bf 537}, 459 (1999); {\bf 602}, 641(E) (2001); A.~Dobado and J.~R.~Pel\'aez, Phys.\ Rev.\ D {\bf 65}, 077502 (2002).

\bibitem{heavyhiggs} O.~Cheyette and M.~K.~Gaillard, Phys.\ Lett.\ B {\bf 197}, 205 (1987); A.~Dobado, M.~J.~Herrero and T.~N.~Truong, Phys.\ Lett.\ B {\bf 235}, 129 (1990). D.~A.~Dicus and W.~W.~Repko, Phys.\ Rev.\ D {\bf 42}, 3660 (1990); Phys.\ Rev.\ D {\bf 44}, 3473 (1991); Phys.\ Rev.\ D {\bf 47}, 4154 (1993); J.~R.~Pel\'aez, Phys.\ Rev.\ D {\bf 55}, 4193 (1997)

\bibitem{full} A.~Denner, S.~Dittmaier and T.~Hahn, Phys.\ Rev.\ D {\bf 56}, 117 (1997); A.~Denner and T.~Hahn, Nucl.\ Phys.\ B {\bf 525}, 27 (1998).

\bibitem{et} J.~M.~Cornwall, D.~N.~Levin, and G.~Tiktopoulos, Phys.\ Rev.\ D {\bf 10}, 1145 (1974); C.~E.~Vayonakis, Lett.\ Nuovo Cim.\  {\bf 17}, 383 (1976).; B.~W.~Lee, C.~Quigg, and H.~B.~Thacker, Phys.\ Rev.\ D {\bf 16}, 1519 (1977); G.~J.~Gounaris, R.~Kogerler, and H.~Neufeld, Phys.\ Rev.\ D {\bf 34}, 3257 (1986); M.~S.~Chanowitz and M.~K.~Gaillard, Nucl.\ Phys.\ {\bf B261}, 379 (1985);
A.~Dobado and J.~R.~Pel\'aez, Nucl.\ Phys.\ {\bf B425}, 110 (1994); Phys.\ Lett.\ {\bf B329}, 469 (1994); 
C.~Grosse-Knetter and I.~Kuss, Z.\ Phys.\ C {\bf 66}, 95 (1995);
H.~J.~He, Y.~P.~Kuang, and X.~Li, Phys.\ Lett.\ {\bf B329}, 278 (1994).

\bibitem{esma} D.~Espriu and J.~Matias, Phys.\ Rev.\ D {\bf 52}, 6530 (1995). 

\bibitem{belen} R.~Alonso, M.~B.~Gavela, L.~Merlo, S.~Rigolin, and J.~Yepes, arXiv:1212.3305v1.

\bibitem{technicolor} A.~Dobado, D.~Espriu, and M.~J.~Herrero, Phys.\ Lett.\ {\bf B255}, 405 (1991);
D.~Espriu and M.~J.~Herrero, Nucl.\ Phys.\ {\bf B373}, 117 (1992);
A.~Dobado, M.~J.~Herrero, J.~R.~Pel\'aez, and E.~Ruiz-Morales, Phys.\ Rev.\ {\bf D} 62, 055011 (2000);

\bibitem{notreeunitarity} B.~Bellazzini, C.~Csaki, J.~Hubisz, J.~Serra and J.~Terning, J.~High Energy Phys. 11 ({\bf 2012}) 003;  Y.~Kikuta and Y.~Yamamoto, arXiv:1210.5674; D.~Choudhury, R.~Islam, A.~Kundu, and B.~Mukhopadhyaya, arXiv:1212.4652; G.~Bhattacharyya, D.~Das, and P.~B.~Pal, Phys.\ Rev.\ D {\bf 87}, 011702 (2013).

\bibitem{bounds} See e.g. G.~F.~Chew, {\it The Analytic S Matrix} (W.A. Benjamin, New York, 1966).

\bibitem{GL} J.~Gasser and H.~Leutwyler, Annals Phys.\  {\bf 158}, 142 (1984); Nucl.\ Phys.\ {\bf B250}, 465 (1985); Nucl.\ Phys.\ {\bf B250}, 517 (1985).

\bibitem{twochannels} T.~Hannah, Phys.\ Rev.\ {\bf D} 55, 5613 (1997).

\bibitem{matching} M.~J.~Herrero and E.~Ruiz-Morales, Nucl.\ Phys.\ {\bf B418}, 431 (1994); Nucl.\ Phys.\ {\bf B437}, 319 (1995); D.~Espriu and J.~Matias, Phys.\ Lett.\ {\bf B341}, 332 (1995);

\bibitem{DW} S.~Dawson and S.~Willenbrock, Phys.\ Rev.\ {\bf D} 40, 2880 (1989).

\bibitem{imaginary} D.~A.~Dicus and W.~W.~Repko, Phys.\ Rev.\ D {\bf 42}, 3660 (1990);  S.~N.~Gupta, J.~M.~Johnson, and W.~W.~Repko, Phys.\ Rev.\ D {\bf 48}, 2083 (1993).

\bibitem{butterworth} J.~M.~Butterworth, B.~E.~Cox, and J.~R.~Forshaw, Phys.\ Rev.\ D {\bf 65}, 096014 (2002).

\bibitem{ewa} G.~L.~Kane, W.~W.~Repko and W.~B.~Rolnick, Phys.\ Lett.\ {\bf B148}, 367 (1984); S.~Dawson, Nucl.\ Phys.\ {\bf B249}, 42 (1985); M.~S.~Chanowitz and M.~K.~Gaillard, Nucl.\ Phys.\ {\bf B261}, 379 (1985).

\bibitem{sixfermion} A.~Alboteanu, W.~Kilian and J.~Reuter, J.~High Energy Phys.~11 ({\bf 2008}) 010; A.~Ballestrero, D.~B.~Franzosi, L.~Oggero and E.~Maina, J.~High Energy Phys.~03 ({\bf 2012}) 031; A.~Ballestrero, D.~B.~Franzosi and E.~Maina, J.~High Energy Phys.~05 ({\bf 2012}) 083;

\bibitem{pdf} J.~Pumplin, D.~R.~Stump, J.~Huston, H.~L.~Lai, P.~M.~Nadolsky and W.~K.~Tung, J.~High Energy Phys.~07 ({\bf 2012}) 012.

\bibitem{tHooft} G.~'t Hooft and M.~J.~G.~Veltman, Nucl.\ Phys.\ {\bf B153}, 365 (1979).

\end{thebibliography}
%%%%%%%%%%%%%%%%s
\newpage

\appendix
%%%%%%%%%%%%%%%%%%%%%%%%%%%%%%%%%%%%%%%%%
\section{NONLINEAR LAGRANGIAN OPERATORS} \label{sec:appendix_operators}
%%%%%%%%%%%%%%%%%%%%%%%%%%%%%%%%%%%%%%%%%
\noindent The full set of $C$, $P$, and $SU(2)_{L} \times U(1)_{Y}$ gauge-invariant $\mathcal{L}_{i}$ are
\be
\begin{array}{ll}
\mathcal{L}_{0} = \frac{1}{4} a_{0} v^{2} T_{\mu} T^{\mu} \hspace{4cm} &  \mathcal{L}_{1} = \frac{1}{2} a_{1} g g' B_{\mu \nu} {\rm Tr} T W^{\mu \nu} \\

\mathcal{L}_{2} = i a_{2} g' B_{\mu \nu} {\rm Tr}\left[ T V^{\mu}V^{\nu} \right] &  \mathcal{L}_{3} = -i a_{3} {\rm Tr}\left[ W_{\mu \nu} \left[V^{\mu},V^{\nu}\right] \right] \\

\mathcal{L}_{4} = a_{4} {\rm Tr}\left[ V_{\mu}V_{\nu} \right]{\rm Tr}\left[ V^{\mu}V^{\nu} \right] & \mathcal{L}_{5} = a_{5} {\rm Tr}\left[ V_{\mu}V^{\mu} \right]{\rm Tr}\left[ V_{\nu}V^{\nu} \right] \\

\mathcal{L}_{6} = a_{6} {\rm Tr}\left[ V_{\mu}V_{\nu} \right] \left( T^{\mu}T^{\nu} \right) & \mathcal{L}_{7} = a_{7} {\rm Tr}\left[ V_{\mu}V^{\mu} \right] \left( T_{\nu}T^{\nu} \right) \\

\mathcal{L}_{8} = - \frac{1}{4} a_{8} g^{2} {\rm Tr}\left[ T W_{\mu \nu} \right]{\rm Tr}\left[ T W^{\mu \nu} \right]  &  \mathcal{L}_{9} = -i a_{9} g {\rm Tr}\left[ T W_{\mu \nu} \right]{\rm Tr}\left[ T V^{\mu}V^{\nu} \right] \\

\mathcal{L}_{10} = a_{10} \left( T_{\mu}T_{\nu} \right) \left( T^{\mu}T^{\nu} \right) & \mathcal{L}_{11} = a_{11}  {\rm Tr}\left[ \left( \mathcal{D}_{\mu}V^{\mu} \right) \left( \mathcal{D}_{\nu}V^{\nu} \right) \right] \\

\mathcal{L}_{12} = a_{12} {\rm Tr}\left[  T \mathcal{D}_{\mu}\mathcal{D}_{\nu}V^{\nu} \right]T^{\mu} & \mathcal{L}_{13} = \frac{1}{2}a_{13}  \left( {\rm Tr}\left[ T \mathcal{D}_{\mu}V_{\nu}\right] \right) \left({\rm Tr}\left[ T \mathcal{D}^{\mu}V^{\nu}\right] \right),

\end{array}
\ee
where
\be
\begin{array}{lll}
V_{\mu} = \left( D_{\mu} U \right) U^{\dagger}  \; , \qquad
&
T = U \tau_{3} U^{\dagger}  \; , \qquad 
& 
T_{\mu} =  {\rm Tr}\left[ T V_{\mu} \right] \\
\multicolumn{3}{l}{\mathcal{D}_{\mu} \mathcal{O}(x) = \partial_{\mu}\mathcal{O}(x) + i g \left[ W_{\mu}, \mathcal{O}(x) \right] \, .}  \\
\end{array}
\ee

The value of the bare coefficients $a_i$ that match the minimal SM Green functions for a heavy Higgs boson at the one-loop 
level are \cite{matching}

\bea
a_{o}^{b} &=& \frac{1}{16\pi^{2}}\frac{3}{8} \left( \Delta_{\epsilon} - \log\fracp{M_{H}^{2}}{\mu^{2}} + \frac{5}{6} \right) \\ \nn
a_{1}^{b} &=& \frac{1}{16\pi^{2}}\frac{1}{12} \left( \Delta_{\epsilon} - \log\fracp{M_{H}^{2}}{\mu^{2}} + \frac{5}{6} \right) \\ \nn
a_{2}^{b} &=& \frac{1}{16\pi^{2}}\frac{1}{24} \left( \Delta_{\epsilon} - \log\fracp{M_{H}^{2}}{\mu^{2}} + \frac{17}{6} \right) \\ \nn
a_{3}^{b} &=& \frac{-1}{16\pi^{2}}\frac{1}{24} \left( \Delta_{\epsilon} - \log\fracp{M_{H}^{2}}{\mu^{2}} + \frac{17}{6} \right) \\ \nn
a_{4}^{b} &=& \frac{-1}{16\pi^{2}}\frac{1}{12} \left( \Delta_{\epsilon} - \log\fracp{M_{H}^{2}}{\mu^{2}} + \frac{17}{6} \right) \\ \nn
a_{5}^{b} &=& \frac{M_{W}^{2}}{2 g^{2} M_{H}^{2}} - \frac{1}{16\pi^{2}}\frac{1}{24} \left( \Delta_{\epsilon} - \log\fracp{M_{H}^{2}}{\mu^{2}} + \frac{79}{3} - \frac{27 \pi}{2 \sqrt{3}} \right) \\ \nn
a_{11}^{b} &=& \frac{-1}{16\pi^{2}}\frac{1}{24} \\ \nn
a_{6}^{b} &=& a_{7}^{b} = a_{8}^{b} = a_{9}^{b} = a_{10}^{b} = a_{12}^{b} = a_{13}^{b} = 0 \, , \\ \nn
\eea

\noindent where $\Delta_{\epsilon} = \left( 2/\epsilon -\gamma_{E} + \log{4\pi} \right)$.  We note that only $a_3$, $a_4$, $a_5$, and $a_{11}$ correspond to custodially symmetric operators, only  $a_{3}$, $a_{4}$, and $a_{5}$ appear in our calculation, and only $a_{4}$ and $a_{5}$ are considered in our numerical results.

%%%%%%%%%%%%%%%%%%%%%%%%%%%%%%%%%
\section{TREE-LEVEL $WW$ SCATTERING AMPLITUDES} \label{sec:appendix_amplitudes}
%%%%%%%%%%%%%%%%%%%%%%%%%%%%%%%%%
\noindent In the isospin limit where $c_{w} \to 1$ ($M_{Z} \to M_{W} \equiv M$) and all $a_{i} \to 0$ except $a_{3}$, $a_{4}$, $a_{5}$, and $a_{11}$, the tree-level and $a_{i}$-dependent amplitude for $\wplus_{L} \wminus_{L} \to Z_{L} Z_{L}$ scattering is given by

\bea
A_{W^{+}W^{-} \to ZZ}^{ {\rm tree} \, + \, a_{i}} \left( p_{1},p_{2},p_{3},p_{4} \right)
& = & 
\;
2 C_{3} \Big[
\eeee{1}{2}{3}{4} 
\Big] \\ \nn
& &
+
C_{4} \Big[
\eeee{1}{4}{2}{3} + \eeee{1}{3}{2}{4} 
\Big]
\\ \nn
& &
- A_{W}^{2} \Bigg\{ \left(\frac{1}{(p_{1}-p_{3})^{2}-M^{2}}\right)
\Bigg[ \\ \nn
& & 
-4 \Big( 
\eepepe{1}{2}{1}{3}{2}{4} + \eepepe{1}{4}{1}{3}{4}{2} + \\ \nn
& & \hspace{0.75cm} 
\eepepe{2}{3}{3}{1}{2}{4} + \eepepe{3}{4}{3}{1}{4}{2}
\Big) \\ \nn
& & 
+2\Big(
\edote{2}{4} \Big(
\pdote{1}{3}(p_{2}+p_{4})\cdot\epsilon_{1} +
\pdote{3}{1}(p_{2}+p_{4})\cdot\epsilon_{3}
\Big) + \\ \nn
& & \hspace{0.75cm} 
\edote{1}{3} \Big(
\pdote{2}{4}(p_{1}+p_{3})\cdot\epsilon_{2} +
\pdote{4}{2}(p_{1}+p_{3})\cdot\epsilon_{4}
\Big) \Big)  \\ \nn
& & \hspace{0.45cm}
- \eeee{1}{3}{2}{4} \Big(
(p_{1}+p_{3})\cdot p_{2} + (p_{2}+p_{4})\cdot p_{1}
\Big)
\Bigg] \\ \nonumber
& &
+ \hspace{0.5cm} (p_{3} \leftrightarrow p_{4})
\Bigg\} 
-g^{2} M^{2} \left(\frac{\edote{1}{2} \edote{3}{4}}{(p_{1}+p_{2})^{2}-M_{H}^{2}} \right) \, , \nn
\eea
where $\epsilon_{i} = \epsilon_{L}(p_{i})$ and the coefficients are given by
\bea
C_{3} & = & g^{2} ( - 1 + g^{2}(a_{5}+2 a_{3}) ) \\ \nn
C_{4} & = & g^{2} (\; \; \: 1 + g^{2}(a_{4}-2 a_{3}) ) \\ \nn
A_{W} & = & -i g (1-g^{2} a_{3}) \, . \ \nn
\eea

%%%%%%%%%%%%%%%%%%%%%%%%%%%%%%%%%
\section{GOLDSTONE-BOSON ONE-LOOP SCATTERING AMPLITUDES} \label{sec:appendix_GBamplitudes}
%%%%%%%%%%%%%%%%%%%%%%%%%%%%%%%%%
\noindent The real parts of the one-loop amplitude for $w^{+} w^{-} \to z z$ scattering are given in Refs.~\cite{DW,imaginary} while the imaginary parts can be found in Ref.~\cite{imaginary}.  We simply reproduce them here, and we will follow closely the notation of Ref.~\cite{DW} after taking $\epsilon \to 0$.  The one-loop amplitude can be written as

\be
\mathcal{M}_{1} = 4 \mathcal{M}_{0}(Z_{w}^{1/2}-1) + \mathcal{M}_{\rm 2-pt} + \mathcal{M}_{\rm 3-pt} + \mathcal{M}_{\rm bubble} +  \mathcal{M}_{\rm tri} +  \mathcal{M}_{\rm box} \, ,
\ee

\noindent where $\mathcal{M}_{0}$ is the tree-level amplitude, given by

\be
\mathcal{M}_{0} = - 2 \lambda \left[ 1 + \frac{m_{H}^{2}}{s-m_{H}^{2}} \right] \, ,
\ee

\noindent with $\lambda \equiv (\frac{1}{8} g^{2} m_{H}^{2}/M_{W}^{2})$.  The wave function renormalization for the vector bosons is given by

\be
Z_{w}^{1/2} = 1 -  \frac{\lambda}{32 \pi^{2}} \, .
\ee

\noindent The two-point corrections to the Higgs propagator are given by

\bea
{\rm Re \,} \mathcal{M}_{\rm 2-pt} &=& - 2 \lambda m_{H}^{2} \frac{ {\rm Re \,}\Pi (s)}{\left(s-m_{H}^{2}\right)^{2}} \\ \nn
{\rm Im \,} \mathcal{M}_{\rm 2-pt} &=& \frac{\lambda^{2}}{\pi} \frac{1}{8} \frac{m_{H}^{4}}{\left( s - m_{H}^{2}\right)^{2}} \left[ 3 \theta(s) + 9 \beta \theta(s-4 m_{H}^{2})  \right]  \, , \\ \nn
\eea

\noindent the three-point corrections to the $h \omega \omega$ vertices are

\bea
{\rm Re \,} \mathcal{M}_{\rm 3-pt} &=& 2 \sqrt{2} \lambda^{1/2} m_{H} \frac{ \Gamma_{3}(m_{H}^{2}/s)}{\left(s-m_{H}^{2}\right)} \\ \nn
{\rm Im \,} \mathcal{M}_{\rm 3-pt} &=& -\frac{\lambda^{2}}{\pi} \frac{1}{4} \frac{m_{H}^{2}}{\left( s - m_{H}^{2}\right)} \left[ -5 \theta(s) + 3 \beta \theta(s-4 m_{H}^{2}) + 2 \frac{m_{h}^{2}}{s} \ln \fracp{s+m_{H}^{2}}{m_{H}^{2}} \theta(s) \right. \\ \nn 
& & \left. \hspace{3cm} + 12 \frac{m_{h}^{2}}{s}\ln \fracp{1+\beta}{1-\beta} \theta(s-4 m_{H}^{2})  \right]  \, ,  \\ \nn
\eea

\noindent and the bubble, triangle, and box diagrams are

\bea
{\rm Re \,} \mathcal{M}_{\rm bubble}  &=& - \frac{\lambda^{2}}{2 \pi^{2}} {\rm Re \,} \left[ \frac{7}{4} \ln\fracp{s}{m_{H}^{2}} + \frac{1}{2} \ln\fracp{t}{m_{H}^{2}} + \frac{1}{2} \ln\fracp{u}{m_{H}^{2}} \right. \\ \nn
& & \hspace{2cm} \left.  + \frac{1}{4} I_{1}\fracp{s}{m_{H}^{2}} + \frac{3}{4} - \frac{9 \pi}{4 \sqrt{3}} \right] \\ \nn
{\rm Im \,} \mathcal{M}_{\rm bubble} &=&  \frac{\lambda^{2}}{\pi} \frac{1}{8} \left[ 7 \theta(s) + 2 \theta(t) + 2 \theta(u) + \beta \theta(s-4 m_{H}^{2})\right] \, , \\ \nn
\eea

\bea
{\rm Re \,} \mathcal{M}_{\rm tri}  &=& - \frac{\lambda^{2}}{2 \pi^{2}}  \left[ G\fracp{m_{H}^{2}}{s} + G\fracp{m_{H}^{2}}{t} + G\fracp{m_{H}^{2}}{u} + H\fracp{m_{H}^{2}}{s}\right]\\ \nn
{\rm Im \,} \mathcal{M}_{\rm tri}  &=& - \frac{\lambda^{2}}{\pi} \frac{1}{2} \left[ \frac{m_{H}^{2}}{s} \ln\fracp{s+m_{H}^{2}}{m_{H}^{2}}\theta(s) +  \frac{m_{H}^{2}}{t} \ln\fracp{t+m_{H}^{2}}{m_{H}^{2}}\theta(t)   \right. \\ \nn
& & \hspace{1.1cm} \left. + \frac{m_{H}^{2}}{u} \ln\fracp{u+m_{H}^{2}}{m_{H}^{2}}\theta(u)  + 2\frac{m_{H}^{2}}{s} \ln\fracp{1+\beta}{1-\beta}\theta(s-4 m_{H}^{2}) \right]  \, , \\ \nn
\eea

\noindent and 

\bea
{\rm Re \,} \mathcal{M}_{\rm box}  &=& - \frac{\lambda^{2}}{4 \pi^{2}}  \left[ F\left(\frac{m_{H}^{2}}{s},\frac{m_{H}^{2}}{t}\right) +  F\left(\frac{m_{H}^{2}}{s},\frac{m_{H}^{2}}{u}\right) \right] \\ \nn
{\rm Im \,} \mathcal{M}_{\rm box}  &=& \left\{ - \frac{\lambda^{2}}{\pi} \frac{1}{2} \frac{m_{H}^{2}}{s} \frac{m_{H}^{2}}{t} \frac{1}{D'} \ln\fracp{D'+\beta}{D'-\beta}\theta(s-4 m_{H}^{2}) + (t \leftrightarrow u) \right. \\ \nn
& & \hspace{0.25cm} \left. - \frac{\lambda^{2}}{\pi} \frac{1}{2} \frac{m_{H}^{2}}{s} \frac{m_{H}^{2}}{t} \frac{1}{D'} \ln\fracp{D'+1}{D'-1}\theta(t) + (t \leftrightarrow u) \right\} \, .  \\ \nn
\eea

\noindent In the above, $\theta(x)$ is the Heaviside step function and

\bea
{\rm Re \,}\Pi (s) &=& \frac{\lambda}{8 \pi^{2}} m_{H}^{2} \left\{ \frac{3}{2} \ln \left|\frac{s}{m_{H}^{2}}\right| + \frac{9}{2} \left[ I_{1}\left( \frac{s}{m_{H}^{2}}\right) - \frac{\pi}{\sqrt{3}} +2 \right] \right\} \\ \nn
\Gamma_{3}(\eta) &=& - \frac{\sqrt{2}}{8 \pi^{2}} \lambda^{3/2} m_{H} \left[ - \frac{5}{2} \ln |\eta| + \frac{21}{4} - \frac{9\pi}{2 \sqrt{3}} + \frac{3}{2} I_{1} \fracp{1}{\eta} + G(\eta) + 3 H(\eta) \right] \\ \nn
I_{1}(a) &=& \left\{
\begin{array}{cc}
2 \fracp{a-4}{a}^{1/2} {\rm arcsinh} \frac{\sqrt{-a}}{2} - 2 &   a<0 \\
2 \fracp{4-a}{a}^{1/2} {\rm arcsin} \frac{\sqrt{a}}{2} - 2 & 0 < a \le 4 \\
2 \fracp{a-4}{a}^{1/2} {\rm arccosh} \frac{\sqrt{a}}{2} - 2 & a>4 \\
\end{array}
\right. \\ \nn
\eea

\noindent and

\bea
F(\sigma,\tau) &=& \sigma \tau \, {\rm Re \,} \left\{ \frac{1}{D} \left[ - {\rm Sp}\fracp{1-\lambda_{+}}{\lambda_{-}} + {\rm Sp}\fracp{-\lambda_{+}}{\lambda_{-}}    \right. \right. \\ \nn
& & \hspace{2.20cm} - {\rm Sp}\fracp{1-\lambda_{+}}{a_{+} - \lambda_{+}} + {\rm Sp}\fracp{-\lambda_{+}}{a_{+} - \lambda_{+}} \\ \nn
& & \hspace{2.20cm} - {\rm Sp}\fracp{1-\lambda_{+}}{a_{-} - \lambda_{+}} + {\rm Sp}\fracp{-\lambda_{+}}{a_{-} - \lambda_{+}} \\ \nn
& & \hspace{2.20cm} \left. \left. +  {\rm Sp}\fracp{1-\lambda_{+}}{- \lambda_{+}} \right] - (\lambda_{+} \leftrightarrow \lambda_{-})\right\} \\ \nn
G(\eta) &=& \eta \, {\rm Re \,} \left[ {\rm Sp}\fracp{1+\eta}{\eta} - \frac{\pi^{2}}{6} \right] \\ \nn
H(\eta) &=& \eta \, {\rm Re \,} \left[ {\rm Sp}\fracp{1-\eta}{x_{+}-\eta} + {\rm Sp}\fracp{1-\eta}{x_{-}-\eta} - {\rm Sp}\fracp{\eta-1}{\eta}  \right. \\ \nn
& & \hspace{0.85cm} \left. -  {\rm Sp}\fracp{-\eta}{x_{+}-\eta} -  {\rm Sp}\fracp{-\eta}{x_{-}-\eta} + \frac{\pi^{2}}{6} \right]  \, , \\ \nn
\eea

\noindent with

\bea
\beta &=& \sqrt{1-\frac{4 m_{H}^{2}}{s}} \\ \nn
D' &=& \sqrt{1-4 \frac{m_{H}^{2}}{s} - 4 \frac{m_{H}^{4}}{s t}} \\ \nn
D  &=& \sqrt{1-4\sigma (1+\tau)}\\ \nn
\lambda_{\pm} &=& \frac{1 \pm D}{2(1+\tau)} \\ \nn
a_{\pm} &=& \frac{1}{2} \left( 1 \pm \sqrt{1-4\sigma} \right) \\ \nn
x_{\pm} &=& \frac{1}{2} \left( 1 \pm \sqrt{1-4\eta} \right)   \, . \\ \nn
\eea

\noindent ${\rm Sp}(z)$ is the Spence function of a complex variable, formally defined as 

\be
{\rm Sp}(z) = - \int_{0}^{z} \frac{dt}{t} \ln (1-t)  \, .
\ee

\noindent It has been approximated in our work using the series expansion described in the appendix of Ref.~\cite{tHooft}, as is also done in Ref.~\cite{DW}.

\end{document}